\newcommand{\typeof}{0}

\documentclass[10pt]{article}

\usepackage{ifthen}
\newcommand{\condinc}[2]{\ifthenelse{\equal{\typeof}{0}}{#1}{#2}}

\condinc{
\newenvironment{varitemize}
{
\begin{list}{\labelitemi}
{\setlength{\itemsep}{0.0mm}
 \setlength{\topsep}{0.0mm}
 \setlength{\parindent}{0.0mm}
 \setlength{\parskip}{0.0mm}
 \setlength{\parsep}{0.0mm}
 \setlength{\partopsep}{0.0mm}
 \setlength{\leftmargin}{\labelwidth}}}
{
 \end{list} 
}}{
\newenvironment{varitemize}
{
\begin{list}{\labelitemii}
{\setlength{\itemsep}{0.0mm}
 \setlength{\topsep}{0.0mm}
 \setlength{\parindent}{0.0mm}
 \setlength{\parskip}{0.0mm}
 \setlength{\parsep}{0.0mm}
 \setlength{\partopsep}{0.0mm}
 \setlength{\leftmargin}{\labelwidth}}}
{
 \end{list} 
}}

\newenvironment{numlist}
{\begin{list}{(\roman{number})}
  {
   \usecounter{number}
   \setlength{\labelwidth}{4.0mm}
   \setlength{\labelsep}{2.0mm}
   \setlength{\leftmargin}{\labelwidth+\labelsep}
   \setlength{\itemindent}{0.0mm}
   \setlength{\listparindent}{\leftmargin+\parindent}
   \setlength{\itemsep}{0.0mm}
   \setlength{\topsep}{0.0mm}
   \setlength{\parskip}{0.0mm}
   \setlength{\parsep}{0.0mm}
   \setlength{\partopsep}{0.0mm}
  }
}
{\end{list}}

\newcommand{\btr}{\!\blacktriangleright\!}
\newcommand{\timearrow}[1]{\stackrel{#1}{\twoheadrightarrow}}
\newcommand{\timearrownp}{\twoheadrightarrow}
\usepackage{calc}
\usepackage{proof}
\usepackage{amsmath,amsfonts,amssymb}
\condinc{\usepackage{a4wide}}{}
\usepackage{stmaryrd}
\usepackage{mathrsfs}
\usepackage{url}

\newcommand{\N}{\mathbb{N}}

\newcommand{\LLL}{{\sf LLL}}
\newcommand{\LAL}{{\sf LAL}}

\newcommand{\MLFPL}{\mathcal{M}}
\newcommand{\MLFPLZ}{\mathcal{M}_0}

\newcommand{\SLL}{{\sf SLL}}
\newcommand{\EAL}{{\sf EAL}}
\newcommand{\SAL}{{\sf SAL}}
\newcommand{\LFPL}{{\sf LFPL}}

\newcommand{\vdashp}[1]{\Vdash_{#1}}
\newcommand{\maj}[4]{#2,#3\vdashp{#1} #4}
\newcommand{\dist}{\mathcal{D}}
\newcommand{\distpar}[3]{\mathcal{D}_{#1}(#2,#3)}
\newcommand{\distpartwo}[4]{\mathcal{D}_{#1}^{#2}(#3,#4)}
\newcommand{\norm}[1]{\mathcal{F}_{#1}}
\newcommand{\normpar}[2]{\mathcal{F}_{#1}(#2)}

\newcommand{\turmac}[2]{\{#1\}(#2)}

\newcommand{\morphism}[4]{#1\stackrel{#2,#3}{\longrightarrow}#4}
\newcommand{\linear}{\multimap}
\newcommand{\tensor}{\otimes}
\newcommand{\timetm}[2]{\mathit{Time}(\turmac{#1}{#2})}

\newcommand{\cp}{\mathit{cp}}
\newcommand{\cons}[2]{\mathit{cons}(#1,#2)}

\newcommand{\emptytree}{\mathit{empty}}
\newcommand{\emptylist}{\mathit{empty}}
\newcommand{\node}[3]{\mathit{node}(#1,#2,#3)}
\newcommand{\basictree}[1]{[#1]_\trees}
\newcommand{\basiclist}[1]{[#1]_\lists}
\newcommand{\basiclistint}[2]{[#1,#2]_\listsint}
\newcommand{\emptytreeone}{\basictree{0}}
\newcommand{\depth}[1]{\mathit{depth}(#1)}

\newcommand{\reasem}[2]{\llbracket #1\rrbracket_{#2}^\mathscr{R}}
\newcommand{\lists}{\mathcal{L}}
\newcommand{\listsint}{\mathcal{I}}
\newcommand{\trees}{\mathcal{T}}

\newcommand{\Listeal}{\mathit{List}_\EAL}
\newcommand{\Listsal}{\mathit{List}_\SAL}
\newcommand{\Listlal}{\mathit{List}_\LAL}
\newcommand{\Hom}[2]{\mathit{Hom}(#1,#2)}

\condinc{
\newcommand{\qed}{\hfill$\Box$}
\newtheorem{theorem}{Theorem}
\newtheorem{proposition}{Proposition}
\newtheorem{lemma}{Lemma}
\newtheorem{definition}{Definition}

\newtheorem{remark}{Remark}
\newtheorem{corollary}{Corollary}
\newenvironment{proof}{\begin{trivlist}
       \item[\hskip \labelsep {\bfseries Proof.}]}{\end{trivlist}}}
{
}
\date{}
\title{Quantitative Models and Implicit Complexity}
\condinc{\author{Ugo Dal Lago\footnote{Dipartimento di Scienze dell'Informazione, 
                 Universit\`a di Bologna. \texttt{dallago@cs.unibo.it}}
                 \and 
                 Martin Hofmann\footnote{Institut f\"ur Informatik,
                 Ludwig-Maximilians-Universit\"at, M\"unchen. \texttt{mhofmann@informatik.uni-muenchen.de}}}}
{\author{Ugo Dal Lago\inst{1}
        \and
        Martin Hofmann\inst{2}}
\institute{{Dipartimento di Scienze dell'Informazione\\ Universit\`a di Bologna\\\texttt{dallago@cs.unibo.it}}
         \and {Institut f\"ur Informatik\\
         Ludwig-Maximilians-Universit\"at, M\"unchen\\\texttt{mhofmann@informatik.uni-muenchen.de}}}}
\begin{document}
\maketitle
\newcounter{number}
\newcounter{numberlist}
\setcounter{numberlist}{1}
\begin{abstract}
  We give new proofs of soundness (all representable functions on base
  types lies in certain complexity classes) for Elementary Affine Logic,
  \LFPL\ (a language for polytime computation close to 
  realistic functional programming introduced by one of us), 
  Light Affine Logic and Soft Affine Logic. The proofs
  are based on a common semantical framework which is merely
  instantiated in four different ways. The framework consists of an
  innovative modification of realizability which allows us to use
  resource-bounded computations as realisers as opposed to including
  all Turing computable functions as is usually the case in
  realizability constructions.
  For example, all realisers in the model for \LFPL\ are polynomially
  bounded computations whence soundness holds by construction of the
  model. The work then lies in being able to interpret all the
  required constructs in the model.
  While being the first entirely semantical proof of polytime
  soundness for light logics, our proof also provides a notable
  simplification of the original already semantical proof of polytime 
  soundness for \LFPL.
  A new result made possible by the semantic framework
  is the addition of polymorphism and a modality to \LFPL\ thus 
  allowing for an internal definition of inductive datatypes.
\end{abstract}
\section{Introduction}
In recent years, a large number of characterizations of
complexity classes based on logics and lambda calculi 
have appeared. At least three different principles have been
exploited, namely linear types~\cite{Bellantoni00apal,hofmann00safe}, 
restricted modalities in the
context of linear logic~\cite{Girard98ic,Asperti02tocl,Lafont04tcs} 
and non-size-increasing computation~\cite{hofmann99lics}.
Although related one to the other, these systems have been
studied with different, often unrelated methodologies and
few results are known about relative intentional expressive
power. We believe that this area of implicit computational
complexity needs unifying frameworks for the analysis of
quantitative properties of computation. This would help
to improve the understanding on existing systems. More
importantly, unifying frameworks can be used \emph{themselves}
as a foundation for controlling the use of resources inside
programming languages.\par
In this paper, we introduce a new semantical framework
which consists of an innovative modification of realizability.
The main idea underlying our proposal lies in considering
bounded-time algorithms as realizers instead of taking plain
Turing Machines as is usually the case
in realizability constructions. Bounds are expressed 
abstractly as elements of a monoid. We can define a model
for a given (logical or type) system by choosing a
monoid flexible enough to justify all the constructs in the system. 
The model can then be used to study the class of representable functions.\par
This allows us to give new proofs of soundness 
(all representable functions on base types lies in certain 
complexity classes) for Light Affine Logic (\LAL, \cite{Asperti02tocl}), 
Elementary Affine Logic (\EAL, \cite{coppola01tlca}), \LFPL~\cite{hofmann99lics} 
and Soft Affine Logic (\SAL, \cite{Baillot04fossacs}). 
While being the first entirely semantical proof of polytime
soundness for light logics, our proof also provides a notable
simplification of the original already semantical proof of polytime
soundness for \LFPL~\cite{hofmann99lics}.
A new result made possible by the semantic framework 
is the addition of polymorphism and a modality to \LFPL.\par
\condinc{
The rest of the paper is organized as follows. In 
Section~\ref{sect:acm} we describe an abstract computational model
that will be used in the rest of the paper. In Section~\ref{sect:ls}
we introduce length spaces and show they can be used to interpret
multiplicative linear logic with free weakening. 
Sections~\ref{sect:els}, \ref{sect:sls} and~\ref{sect:lls} are devoted
to present instances of the framework together with soundness results for
elementary, soft and light affine logics. Section~\ref{sect:lfpl}
presents a further specialization of length spaces and a new
soundness theorem for \LFPL\ based on it.\par}
{The rest of the paper is organized as follows. This
Section is devoted to a brief description of related work and
to preliminaries. In Section~\ref{sect:ls}
we introduce length spaces and show they can be used to interpret
multiplicative linear logic with free weakening. 
Sections~\ref{sect:els} and~\ref{sect:oll} are devoted
to present instances of the framework together with soundness results for
elementary, soft and light affine logics. Section~\ref{sect:lfpl}
presents a further specialization of length spaces and a new
soundness theorem for \LFPL\ based on it.\par}
\condinc{}{An extended version of this paper is available~\cite{dallago05}.}

\paragraph{Related-Work}
Realizability has been used in connection with resource-bounded
computation in several places. The most prominent is
Cook and Urquhart work~\cite{Cook93apal}, where terms of a language called $\textit{PV}^\omega$ are 
used to realize formulas of bounded arithmetic. The contribution of
that paper is related to ours in that realizability is used to show
``polytime soundness'' of a logic. There are important differences
though. First, realizers in Cook and Urquhart~\cite{Cook93apal} 
are typed and very closely related to the logic that is being realized. Second, the
language of realizers $\textit{PV}^\omega$  only contains first order recursion
and is therefore useless for systems like \LFPL\ or \LAL. In contrast,
we use untyped realizers and interpret types as certain partial
equivalence relations on those. This links our work to the untyped 
realizability model HEO (due to Kreisel~\cite{Kreisel59}). This, in turn,
has also been done by Crossley et al.~\cite{Crossley94jmlcs}. There, however, one proves externally
that untyped realizers (in this case of bounded arithmetic formulas)
are polytime. In our work, and this happens for the first time, the
untyped realizers are used to give meaning to the logic and obtain
polytime soundness as a corollary. Thus, certain resource bounds
are built into the untyped realizers by their very construction.
Such a thing is not at all obvious, because untyped universes of
realizers tend to be Turing complete from the beginning to due
definability of fixed-point combinators. We get around this problem
through our notion of a resource monoid and addition of a certain
time bound to Kleene applications of realizers. Indeed, we consider
this as the main innovation of our paper and hope it to be useful
elsewhere.
\condinc{
\section{A Computational Model}\label{sect:acm}
In this paper, we rely on an abstract computational framework rather
than a concrete one like Turing Machines. This, in particular, will simplify
proofs.\par 
Let $L\subseteq\Sigma^*$ be the set of finite sequences over the
alphabet $\Sigma$. We assume a
pairing function $\langle\cdot,\cdot\rangle :
L\times L\rightarrow L$ and a length function
$|\cdot| : L\rightarrow \N$ such that $|\langle x,y\rangle| =
|x|+|y|+\cp$ and $|x|\leq \textit{length}(x)$, where
$\textit{length}(x)$ is the number of symbols in $x$ and $\cp$ is
a fixed constant. We assume a reasonable encoding of
algorithms as elements of $L$. We write
$\turmac{e}{x}$ for the (possibly undefined) application
of algorithm $e\in L$ to input $x\in L$. We furthermore assume
an abstract time measure $\timetm{e}{x}\in\N$ such that $\timetm{e}{x}$ 
is defined whenever $\turmac{e}{x}$ is and, moreover
\begin{varitemize}
\item $\turmac{e}{x}$ can be evaluated on a Turing machine in time
  bounded by $p(\timetm{e}{x}+|e|+|x|)$, where $p:\N\rightarrow\N$ is
  a fixed polynomial.
\item For each Turing machine $M$ running in time $f:\N\rightarrow\N$,
  there is $e\in L$ so that $\turmac{e}{\Phi(x)}=\Phi(y)$, 
  (where $y$ is the result of running $M$ on input
  $x$). Furthermore, $\timetm{e}{\Phi(x)}=O(f(|x|))$.
\item $B=\{0,1\}^*$ can be embedded into $L$ by a map $\Phi:B\rightarrow L$
  such that both $\Phi$ and $\Phi^{-1}$ can be computed in 
  polynomial time.
\item There are $e_0,e_1\in L$ such that for every $x\in B$,
  $\turmac{e_0}{\Phi(x)}=\Phi(0x)$, $\turmac{e_1}{\Phi(x)}=\Phi(1x)$.
  Moreover, $\timetm{e_0}{x}=\timetm{e_1}{x}=O(1)$.
\item There is $e_\mathit{comp}$ (composition) such that for every $x,y$
  it holds that $\turmac{e_\mathit{comp}}{\langle x,y\rangle}=z$ where
  $|z| = |x|+|y|+O(1)$ and $\turmac{z}{w}=\turmac{y}{\turmac{x}{w}}$;
  moreover, $\timetm{e_\mathit{comp}}{\langle x,y\rangle}=O(1)$ 
  and $\timetm{e_\mathit{comp}}{w} = \timetm{x}{w}+\timetm{y}{\turmac{x}{w}}+O(1)$.
\item There is $e_\mathit{id}$ (identity) such that
  $\turmac{e_\mathit{id}}{x}=x$ for every $x$ and
  $\timetm{e_\mathit{id}}{x}= O(1)$.
\item For every $x\in L$ there is $e_\mathit{const}^x$ such
  that $\turmac{e_\mathit{const}^x}{y}=x$ and $\timetm{e_\mathit{const}^x}{y}=O(1)$.
\item For every $x\in L$ there is $e_\mathit{tensconst}^x$ such
  that $\turmac{e_\mathit{tensconst}^x}{y}=\langle y,x\rangle$ and
  $\timetm{e_\mathit{tensconst}^x}{y}=O(1)$.
\item There is $e_\mathit{throwfirst}$ such that for every $x\in L$
  $\turmac{e_\mathit{throwfirst}}{\langle x,y\rangle}=y$ and
  $\timetm{e_\mathit{throwfirst}}{\langle x,y\rangle}=O(1)$.
\item There is $e_\mathit{swap}$ (swapping) such that 
  $\turmac{e_\mathit{swap}}{\langle x,y\rangle}=\langle
  y,x\rangle$ and $\timetm{e_\mathit{swap}}{z} \leq O(1)$.
\item There is $e_\mathit{tens}$ (tensor) such that
  for every $x$ $\turmac{e_\mathit{tens}}{x}=y$ where
  $|y|=|x|+O(1)$ and $\turmac{y}{\langle
  z,w\rangle}=\langle\turmac{x}{z},w\rangle$; moroever,
  $\timetm{e_\mathit{tens}}{x}=O(1)$ and 
  $\timetm{y}{\langle z,w\rangle}=\timetm{x}{z}+O(1)$.
\item There is $e_{\mathit{assl}}$ (rebracketing) such that
  $\turmac{e_{\mathit{assl}}}{\langle x,\langle
  y,z\rangle\rangle}=\langle\langle x,y\rangle,z\rangle$ and
  $\timetm{e_{\mathit{assl}}}{x}=O(1)$. 
\item There is $e_\mathit{contr}$ (duplication, copying) such that
  $\turmac{e_\mathit{contr}}{x}=\langle x,x\rangle$ and
  $\timetm{e_{\mathit{contr}}}{x} = O(|x|)$. 
\item There is $e_\mathit{eval}$ (application) such that
  $\turmac{e_\mathit{eval}}{\langle x,y\rangle}=\turmac{x}{y}$ and
  $\timetm{e_\mathit{eval}}{\langle
    x,y\rangle}=\timetm{x}{y}+O(1)$. 
\item There is $e_\mathit{curry}$ (currying, ``smn-theorem'') 
  such that, for each $x$, $y=\turmac{e_\mathit{curry}}{x}$ 
  exists and satisfies $|y|=|x|+O(1)$ and $\timetm{e_\mathit{curry}}{x}=O(1)$; moreover,
  for every $z$, $c_z=\turmac{y}{z}$ exists and satisfies $|c_z|=|y|+|z|+O(1)$ and
  $\timetm{y}{z}=O(1)$; finally, for every $w$, $\turmac{c_z}{w}=\turmac{x}{\langle z,w\rangle}$ 
  and $\timetm{c_z}{w}=\timetm{x}{\langle z,w\rangle}+O(1)$.  
\end{varitemize}
There are a number of ways to instantiate this framework. One noticeable
and simple way consists in using call-by-value lambda calculus and is 
described in the following. $\Sigma$ will be $\{\lambda,@,0,1,\btr\}$. 
To any lambda term $M\in\Lambda$, we can associate a string $M^\#\in\Sigma^*$
in the obvious way. For example, if $M\equiv (\lambda x.xy)(\lambda x.\lambda y.\lambda z.x)$,
then $M^\#$ is
$$
@\lambda @\btr 0\btr\lambda\lambda\lambda\btr 1 0
$$
In other words, free occurrences of variables are translated into $\btr$, while
bounded occurrences of variables are translated into $\btr s$, where $s$ is the
binary representation of the deBruijn index for the occurrence. $L$ will
just be the set of strings in $\Sigma^*$ corresponding to
lambda terms via the mapping we just described. In the following,
we will often write a lambda-term in the usual notation, but this is
just syntactic sugar for the corresponding element of $L$. The abstract 
length $|s|$ of $s\in\ L$ is just $\mathit{length}(s)$.
The map $\Phi:B\rightarrow L$ is defined by induction as follows:
\begin{eqnarray*}
  \Phi(\varepsilon)&=&\lambda x.\lambda y.\lambda z.z\\
  \Phi(0s)&=&\lambda x.\lambda y.\lambda z.x\Phi(s)\\
  \Phi(1s)&=&\lambda x.\lambda y.\lambda z.y\Phi(s)
\end{eqnarray*}
Given $M,N\in\Lambda$, consider the following definitions:
\begin{eqnarray*}
  \langle M,N\rangle &\equiv& \lambda x.xMN\\
  M_0&\equiv&\lambda x.\lambda y.\lambda z.\lambda w.yx\\
  M_1&\equiv&\lambda x.\lambda y.\lambda z.\lambda w.zx\\
  M_\mathit{comp}&\equiv&\lambda x.\lambda y.\lambda z.x(yz)\\
  M_\mathit{id}&\equiv&\lambda x.x\\
  M_\mathit{const}^N&\equiv&\lambda x.N\\
  M_\mathit{tensconst}^N&\equiv&\lambda x.\lambda y.yxM\\
  M_\mathit{throwfirst}&\equiv&\lambda x.x(\lambda y.\lambda z.z)\\
  M_\mathit{swap}&\equiv&\lambda x.x(\lambda y.\lambda w.\lambda z.zwy)\\
  M_\mathit{tens}&\equiv& \lambda x.\lambda y.y(\lambda z.\lambda q.(\lambda y.\lambda w.wyq)(xz))\\
  M_\mathit{assl}&\equiv&\lambda x.x(\lambda y.\lambda w.w(\lambda z.\lambda q.\lambda r.r(\lambda s.syz)q))\\
  M_\mathit{contr}&\equiv&\lambda x.\lambda y.yxx\\
  M_\mathit{eval}&\equiv&\lambda x.x(\lambda y.\lambda w.yw)\\
  M_\mathit{curry}&\equiv&\lambda x.\lambda y.\lambda w.x(\lambda z.zyw)
\end{eqnarray*}
\emph{Values} are abstractions and variables.
We consider call-by-value reduction on lambda terms, i.e. we take $\rightarrow$
as the closurure of
$$
(\lambda x.M)V\rightarrow M\{x/V\}
$$
under all applicative contexts.
The application $\turmac{M}{N}$ of two lambda terms is the normal
form of $MN$ relative to the call-by-value reduction (if one exists).
We now define a (ternary) relation 
$\timearrownp\;\subseteq\Lambda\times\N\times\Lambda$. 
In the following, we will write $M\timearrow{n}N$ standing
for $(M,n,N)\in\timearrownp$ The precise definition
of $\timearrownp$ (in SOS-style) follows:
$$
\begin{array}{ccccc}
\infer{M\timearrow{0}M}{} &\hspace{1cm}&
\infer{M\timearrow{n}N}{M\rightarrow N & n=\max\{1,|N|-|M|\}} &\hspace{1cm}&
\infer{M\timearrow{n+m}L}{M\timearrow{n}N & N\timearrow{m}L}
\end{array}
$$
It turns out that for every $M,N$ such that $L$ is the normal form of $MN$, 
there is exactly one integer $n$ such that $MN\timearrow{n}L$. So, defining
$\timetm{M}{N}$ to be just $n$ is unambiguous. All the axioms listed at the
beginning of this section can be proved to be satisfied by this calculus. 
}{
\paragraph{Preliminaries}
In this paper, we rely on an abstract computational framework rather
than a concrete one like Turing Machines. This, in particular, will simplify
proofs.\par 
Let $L\subseteq\Sigma^*$ be the set of finite sequences over the
alphabet $\Sigma$. We assume a
pairing function $\langle\cdot,\cdot\rangle :
L\times L\rightarrow L$ and a length function
$|\cdot| : L\rightarrow \N$ such that $|\langle x,y\rangle| =
|x|+|y|+\cp$ and $|x|\leq \textit{length}(x)$, where
$\textit{length}(x)$ is the number of symbols in $x$ and $\cp$ is
a fixed constant. We assume a reasonable encoding of
algorithms as elements of $L$. We write
$\turmac{e}{x}$ for the (possibly undefined) application
of algorithm $e\in L$ to input $x\in L$. We furthermore assume
an abstract time measure $\timetm{e}{x}\in\N$ such that $\timetm{e}{x}$ 
is defined whenever $\turmac{e}{x}$ is and, moreover, there exists a fixed
polynomial $p$ such that $\turmac{e}{x}$ can be evaluated on a Turing machine in time
bounded by $p(\timetm{e}{x}+|e|+|x|)$ (this is related to the so-called
invariance thesis~\cite{Boas90}). By ``reasonable'', we mean for
example that for any $e,d\in L$ there exists $d\circ e\in L$  such that
$|d\circ e| = |d|+|e|+O(1)$ and $\turmac{d\circ e}{x}=\turmac{d}{y}$ where 
$y=\turmac{e}{x}$ and moreover $\timetm{d\circ e}{x} = \timetm{e}{x}+
\timetm{d}{y}+O(1)$. We furthermore assume that the abstract time needed to compute $d\circ e$ 
from $\langle d,e\rangle$ is constant. Likewise, we assume that
``currying'' and rewiring operations such as $\langle x,\langle
y,z\rangle\rangle\mapsto \langle\langle y,z\rangle x\rangle$
can be done in constant time. However, we do allow linear 
(in $|x|$) abstract time for copying operations such us 
$x\mapsto \langle x,x\rangle$.\par
There are a number of ways to instantiate this framework. In the
appendix, the precise form of the assumptions we make as well as 
one instance based on call-by-value lambda-calculus are briefly 
described.}
\section{Length Spaces}\label{sect:ls}
In this section, we introduce the category of length spaces and study
its properties. Lengths will not necessarily be numbers but rather 
elements of a commutative monoid.\par
A \emph{resource monoid} is a quadruple $M=(|M|,+,\leq_M,\dist_M)$ where
\begin{numlist}
  \item
  $(|M|,+)$ is a commutative monoid;
  \item
  $\leq_M$ is a pre-order on $|M|$ which is compatible with $+$;
  \item
  $\dist_M:\{(\alpha,\beta)\;|\;\alpha\leq_M\beta\}\rightarrow\N$ is a function such
  that for every $\alpha,\beta,\gamma$
  \begin{eqnarray*}
  \distpar{M}{\alpha}{\beta}+\distpar{M}{\beta}{\gamma}&\leq&\distpar{M}{\alpha}{\gamma}\\
  \distpar{M}{\alpha}{\beta}&\leq&\distpar{M}{\alpha+\gamma}{\beta+\gamma}
  \end{eqnarray*}
  and, moreover, for every $n\in\N$ there is $\alpha$ such
  that $\distpar{M}{0}{\alpha}\geq n$.
\end{numlist}
Given a resource monoid $M=(|M|,+,\leq_M,\dist_M)$, the function
$\norm{M}:|M|\rightarrow\N$ is defined by putting
$\normpar{M}{\alpha}=\distpar{M}{0}{\alpha}$. We abbreviate 
$\sigma+\dots+\sigma$ ($n$ times) as $n.\sigma$.\par 
Let us try to give some intuition about these axioms. We shall use
elements of a resource monoid to bound data, algorithms, and runtimes
in the following way: an element $\varphi$ bounds an algorithm $e$ if
$\normpar{M}{\varphi}\geq |e|$ and, more importantly, whenever $\alpha$
bounds an input $x$ to $e$ then there must be a bound
$\beta\leq_M\varphi+\alpha$ for the result $y=\turmac{e}{x}$ and, most
importantly, the runtime of that computation must be bounded by
$\distpar{M}{\beta}{\varphi+\alpha}$. So, in a sense, we have the option
of either producing a large output fast or to take a long time for a
small output. The ``inverse triangular'' law above ensures that the
composition of two algorithms bounded by $\varphi_1$ and $\varphi_2$,
respectively, can be bounded by $\varphi_1+\varphi_2$ or a simple
modification thereof. In particular, the contribution of the
unknown intermediate result in a composition cancels out using
that law. Another useful intuition is that
$\distpar{M}{\alpha}{\beta}$ behaves like the difference
$\beta-\alpha$, indeed, $(\beta-\alpha)+(\gamma-\beta)\leq
\gamma-\alpha$.\par
\condinc{
\begin{lemma}
If $M$ is a resource monoid, then $\dist_M$ is antitone on its first argument and 
monotone on its second argument.
\end{lemma}
\begin{proof}
If $\alpha\leq_M\beta$, then
\begin{eqnarray*}
\distpar{M}{\alpha}{\gamma}&\geq&\distpar{M}{\alpha}{\beta}+\distpar{M}{\beta}{\gamma}
                          \geq\distpar{M}{\beta}{\gamma};\\
\distpar{M}{\gamma}{\alpha}&\leq&\distpar{M}{\gamma}{\alpha}+\distpar{M}{\alpha}{\beta}
                          \geq\distpar{M}{\gamma}{\beta}.
\end{eqnarray*}
This concludes the proof.\qed
\end{proof}}{}
A \emph{length space} on a resource monoid $M=(|M|,+,\leq_M,\dist_M)$
is a pair $A=(|A|,\vdashp{A})$, where $|A|$ is a set
and $\vdashp{A}\;\subseteq|M|\times L\times |A|$
is a (infix) relation satisfying the following conditions:
\begin{numlist}
  \item
  If $\maj{A}{\alpha}{e}{a}$, then $\normpar{M}{\alpha}\geq |e|$;
  \item
  For every $a\in|A|$, there are $\alpha,e$ such that $\maj{A}{\alpha}{e}{a}$
  \item
  If $\maj{A}{\alpha}{e}{a}$ and $\alpha\leq_M\beta$, then
  $\maj{A}{\beta}{e}{a}$;
  \item
  If $\maj{A}{\alpha}{e}{a}$ and $\maj{A}{\alpha}{e}{b}$, then $a=b$.
\end{numlist}
The last requirement implies that each element of $|A|$ is uniquely
determined by the (nonempty) set of it realisers and in particular
limits the cardinality of any length space to the number of partial
equivalence relations on $L$.\par
A \emph{morphism} from length space $A=(|A|,\vdashp{A})$ to length space
$B=(|B|,\vdashp{B})$ (on the same resource monoid $M=(|M|,+,\leq_M,\dist_M)$)
is a function  $f:|A|\rightarrow |B|$ such that there exist 
 $e\in L=\Sigma^*$, $\varphi\in |M|$ with $\normpar{M}{\varphi}\geq |e|$ and
whenever $\maj{A}{\alpha}{d}{a}$, there must be $\beta,c$ such that 
\begin{numlist}
  \item
  $\maj{B}{\beta}{c}{f(a)}$;
  \item
  $\beta\leq_M\varphi+\alpha$;
  \item
  $\turmac{e}{d}=c$;
  \item
  $\timetm{e}{d}\leq\distpar{M}{\beta}{\varphi+\alpha}$
\end{numlist}
We call $e$ a realizer of $f$ and $\varphi$ a majorizer of $f$.
The set of all morphisms from $A$ to $B$ is denoted as $\Hom{A}{B}$.
If $f$ is a morphism from $A$ to $B$ realized by $e$ and majorized by
$\varphi$, then we will write
$f:\morphism{A}{e}{\varphi}{B}$ or $\maj{A\linear B}{\varphi}{e}{f}$. 
\begin{remark}\label{remark:oldstuff}
It is possible to alter the time bound in the definition
of a morphism to 
$\timetm{e}{d}\leq\distpar{M}{\beta}{\varphi+\alpha}\normpar{M}{\alpha+\varphi}$.
This allows one to accommodate linear time operations by padding
the majorizer for the morphism. All the subsequent proofs
go through with this alternative definition, at the expense of 
simplicity and ease of presentation,
\end{remark}\par
Given two length spaces $A=(|A|,\vdashp{A})$ and $B=(|B|,\vdashp{B})$ on the
same resource monoid $M$, we can build
$A\otimes B=(|A|\times |B|,\vdashp{A\otimes B})$ (on $M$)
where $e,\alpha\vdashp{A\otimes B}(a,b)$ iff $\normpar{M}{\alpha}\geq |e|$ and 
there are $f,g,\beta,\gamma$ with 
$$
\begin{array}{c}
f,\beta\vdashp{A} a\\
g,\gamma\vdashp{B} b\\
e=\langle f,g\rangle\\
\alpha\geq_M\beta+\gamma
\end{array}
$$
$A\otimes B$ is a well-defined length space due to the axioms on $M$.\par
Given $A$ and $B$ as above, we can build $A\linear B=(\Hom{A}{B},\vdashp{A\linear B})$
where $e,\alpha\vdashp{A\linear B}f$ iff $f$ is a morphism from $A$
to $B$ realized by $e$ and majorized by $\alpha$.\par
\condinc{
Morphisms can be composed:
\begin{lemma}[Composition]
Given length spaces $A,B,C$, there is a morphism 
$$
\mathit{comp}:(B\linear C)\otimes(A\linear B)\rightarrow(A\linear C)
$$ 
such that $\mathit{comp}(f,g)=\lambda x.f(g(x))$.
\end{lemma}
\begin{proof}
Let $f:\morphism{A}{x}{\varphi}{B}$ and $g:\morphism{B}{y}{\psi}{C}$. 
We know there are constants $p,q,r$ such
that $\turmac{e_\mathit{comp}}{\langle x,y\rangle}=z$ where
$|z|\leq |x|+|y|+p$ and $\turmac{z}{w}=\turmac{y}{\turmac{x}{w}}$;
moreover, $\timetm{e_\mathit{comp}}{\langle x,y\rangle}\leq r$ 
and $\timetm{e_\mathit{comp}}{w}=\timetm{x}{w}+\timetm{y}{\turmac{x}{w}}+q$. Now,
let us now choose $\mu$ such that $\normpar{M}{\mu}\geq p+q$,
We will prove that $comp(f,g):\morphism{A}{z}{\varphi+\psi+\mu}{C}$.
Obviously, $\normpar{M}{\varphi+\psi+\mu}\geq |z|$. If $\maj{A}{\alpha}{w}{a}$,
then there must be $\beta,t$ such that $\maj{B}{\beta}{t}{f(a)}$ 
and the other conditions prescribed by the definition of a morphism 
hold. Moreover, there must be $\gamma,s$ such 
that $\maj{C}{\gamma}{s}{g(f(a))}$ and, again, the other conditions
are satisfied. Putting them together, we get:
$$ 
\gamma\leq_M\beta+\psi\leq_M\alpha+\varphi+\psi\leq_M\alpha+\varphi+\psi+\mu
$$ 
and 
\begin{eqnarray*}
\timetm{z}{w}&\leq&\timetm{x}{w}+\timetm{y}{t}+q\\
             &\leq&\distpar{M}{\beta}{\alpha+\varphi}+
                   \distpar{M}{\gamma}{\beta+\psi}+\normpar{M}{\mu}\\
             &\leq&\distpar{M}{\beta+\psi}{\alpha+\varphi+\psi}+
                   \distpar{M}{\gamma}{\beta+\psi}+\distpar{M}{0}{\mu}\\
             &\leq&\distpar{M}{\gamma}{\alpha+\varphi+\psi+\mu}
\end{eqnarray*}
This concludes the proof, since 
$comp:\morphism{(B\linear C)\otimes(A\linear B)}{(f,g)}{\xi}{A\linear C}$
where $\xi$ is such that $\normpar{M}{\xi}\geq r+|e_\mathit{comp}|$.\hfill $\Box$
\end{proof}
Basic morphisms can be built independently on the underlying resource monoid. Noticeably,
they correspond to axiom of multiplicative linear logic:
\begin{lemma}[Basic Maps]
Given length spaces $A,B,C$, there are morphisms:
\begin{eqnarray*}
\mathit{id}&:&A\rightarrow A\\
\mathit{swap}&:&A\otimes B\rightarrow B\otimes A\\
\mathit{assl}&:&A\otimes(B\otimes C)\rightarrow (A\otimes B)\otimes C\\
\mathit{eval}&:&A\otimes(A\linear B)\rightarrow B\\
\mathit{curry}&:&((A\otimes B)\linear C)\rightarrow A\linear (B\linear C)
\end{eqnarray*}
where
\begin{eqnarray*}
\mathit{id}(a)&=&a\\
\mathit{swap}(a,b)&=&(b,a)\\
\mathit{assl}(a,(b,c))&=&((a,b),c)\\
\mathit{eval}(a,f)&=&f(a)\\
\mathit{curry}(f)&=&\lambda a.\lambda b.f(a,b)
\end{eqnarray*}
\end{lemma}
\begin{proof}
We know that $\{e_\mathit{id}\}(d)$ takes constant time,
say at most $p$. Then, let $\varphi_\mathit{id}\in M$ be
such that $\normpar{M}{\varphi_\mathit{id}}\geq p+|e_\mathit{id}|$ 
(this can always be done). Now, let $\maj{A}{\alpha}{d}{a}$. We have
that $\maj{A}{\alpha}{d}{\mathit{id}(a)}$, $\alpha\leq_M\alpha+\varphi_\mathit{id}$,
$\{e_\mathit{id}\}(d)=d$. Moreover
\begin{eqnarray*}
\timetm{e_\mathit{id}}{d}&\leq&p\leq\normpar{M}{\varphi_\mathit{id}}=\distpar{M}{0}{\varphi_\mathit{id}}\\
&\leq& \distpar{M}{\alpha}{\alpha+\varphi_\mathit{id}}
\end{eqnarray*}
This proves $\mathit{id}$ to be a morphism.\par
We know that $\{e_\mathit{swap}\}(\langle d,c\rangle)$
takes constant time, say at most $p$. Then, 
let $\varphi_\mathit{swap}\in |M|$ be
such that $\normpar{M}{\varphi_\mathit{id}}\geq p+|e_\mathit{swap}|$.
Now, let $\maj{A\otimes B}{\alpha}{e}{(a,b)}$. This i that
$e=\langle d,c\rangle$ and $\maj{B\otimes A}{\alpha}{\langle c,d\rangle}{(b,a)}$.
We can then apply the same argument as for $\mathit{id}$. In particular:
\begin{eqnarray*}
\timetm{e_\mathit{swap}}{e}&\leq&p\leq\normpar{M}{\varphi_\mathit{swap}}=
  \distpar{M}{0}{\varphi_\mathit{swap}}\\
&\leq& \distpar{M}{\alpha}{\alpha+\varphi_\mathit{swap}}
\end{eqnarray*}
This proves $\mathit{swap}$ to be a morphism. We can verify $\mathit{assl}$  to be a morphism exactly in the same way.\par
We know that
$\{e_\mathit{eval}\}(\langle d,c\rangle)=\{d\}(c)$ and
$\{e_\mathit{eval}\}(\langle d,c\rangle)$ takes constant overload time, say at most 
$p$. $\varphi_\mathit{eval}$ is chosen as to satisfy 
$\normpar{M}{\varphi_{\mathit{eval}}}\geq p$.
Let now $\maj{A\otimes (A\linear B)}{\alpha}{e}{(a,f)}$. This means that
$e=\langle d,c\rangle$ and there are $\beta$ and $\gamma$ such that
$$
\begin{array}{c}
\maj{A}{\beta}{d}{a}\\
\maj{A\linear B}{\gamma}{c}{f}\\
\alpha\geq_M\beta+\gamma\\
\normpar{M}{\alpha}\geq\normpar{M}{\beta}+\normpar{M}{\gamma}+\cp
\end{array}
$$
From $\maj{A\linear B}{\gamma}{c}{f}$ it follows
that, by the definition of a morphism,
there must be $\delta,h$ such that
\begin{numlist}
  \item
  $\maj{B}{\delta}{h}{f(a)}$
  \item
  $\delta\leq_M\beta+\gamma$
  \item
  $\{c\}(d)=h$
  \item
  $\timetm{c}{d}\leq\distpar{M}{\delta}{\beta+\gamma}$
\end{numlist}
From $\delta\leq_M\beta+\gamma$ and $\beta+\gamma\leq_M\alpha$, it
follows that $\delta\leq_M\alpha\leq_M\alpha+\mu$. 
Moreover:
\begin{eqnarray*}
\timetm{e_\mathit{eval}}{\langle d,c\rangle}&\leq&p+\timetm{c}{d}
  \leq\normpar{M}{\varphi_\mathit{eval}}+\distpar{M}{\delta}{\beta+\gamma} \\
&\leq&\normpar{M}{\varphi_\mathit{eval}}+\distpar{M}{\delta}{\beta+\gamma}
  +\distpar{M}{\beta+\gamma}{\alpha} \\
&\leq&\distpar{M}{0}{\varphi_\mathit{eval}}+\distpar{M}{\delta}{\alpha} \\
&\leq&\distpar{M}{\delta}{\alpha+\varphi_\mathit{eval}}
\end{eqnarray*}
Now, let us prove that $\mathit{curry}$ is a morphism.
First of all, we know there must be constants $p,q,r,s,t$ such
that, for each $e,x,y$, there are $d$ and $c_x$ with
\begin{eqnarray*}
\timetm{e_\mathit{curry}}{e}&\leq&p\\
d&=&\turmac{e_\mathit{curry}}{e}\\
|d|&\leq&|e|+q\\
\timetm{d}{x}&\leq&r\\
c_x&=&\turmac{d}{x}\\
|c_x|&\leq&|e|+|x|+s\\
\timetm{c_x}{y}&\leq&\timetm{e}{\langle x,y\rangle}+t\\
\turmac{e}{\langle x,y\rangle}&=&\turmac{c_x}{y}
\end{eqnarray*}
Let $\mu,\theta,\xi\in |M|$ be such that
\begin{eqnarray*}
\normpar{M}{\xi}&\geq& p\\
\normpar{M}{\mu}&\geq& q\\
\normpar{M}{\sigma}&\geq& r\\
\normpar{M}{\theta}&\geq& s\\
\normpar{M}{\eta}&\geq& t\\
\normpar{M}{\chi}&\geq& \cp
\end{eqnarray*}
Let now $\maj{A\otimes B\linear C}{\gamma}{e}{f}$.
We know that $|d|\leq|e|+q$ and
$\timetm{e_\mathit{curry}}{e}\leq p$. In order
to prove that $\mathit{curry}$ is indeed a morphism
realized by $e_\mathit{curry}$ and majorized by
$\mu+\xi+\sigma+\theta+\chi+\eta$, it then suffices
to prove that 
$$
\maj{A\linear B\linear C}{\gamma+\mu+\sigma+\theta+\chi+\theta}{d}{\lambda a.\lambda b.f(a,b)}.
$$
Let then $\maj{A}{\alpha}{x}{a}$. There is
$c_x$ such that $c_x=\turmac{d}{x}$, 
$|c_x|\leq|e|+|x|+s$ and $\timetm{d}{x}\leq r$. In
order to prove that $\lambda a.\lambda b.f(a,b)$ is
indeed a morphism realized by $d$ and majorized 
by $\gamma+\mu+\sigma+\theta+\chi+\eta$, it then suffices to prove that
$\maj{B\linear C}{\gamma+\alpha+\mu+theta+\chi+\eta}{c_x}{\lambda b.f(a,b)}$.
Let then $\maj{B}{\beta}{y}{b}$. There are $\delta,c$ such
$\maj{C}{\delta}{c}{f(a,b)}$, where 
$\delta\leq\alpha+\beta+\chi+\gamma$. Moreover, 
we know that
\begin{eqnarray*}
\timetm{c_x}{y}&\leq&\timetm{e}{\langle x,y\rangle}+t\leq\distpar{M}{\delta}{\alpha+\beta+\chi+\gamma}+t\\
&\leq&\distpar{M}{\delta}{\alpha+\beta+\gamma+\chi}+\distpar{M}{0}{\eta+\mu+\theta}\\
&\leq&\distpar{M}{\delta}{\alpha+\beta+\gamma+\chi+\eta+\mu+\theta}
\end{eqnarray*}
This concludes the proof.\qed.
\end{proof}
Length spaces can justify the usual rule for tensor as a map-former:
\begin{lemma}[Tensor]
Given length spaces $A,B,C$, there is a morphism
$$
\mathit{tens}:(A\linear B)\rightarrow((A\otimes C)\linear(B\otimes C))
$$
where $\mathit{tens}(f)=\lambda x.(f(\pi_1(x)),\pi_2(x))$.
\end{lemma}
\begin{proof}
Let $f:\morphism{A}{x}{\varphi}{B}$. We know there are constants
$p,q$ such that
$\turmac{e_\mathit{tens}}{x}=y$ where
$|y|\leq|x|+p$ and $\turmac{y}{\langle
z,w\rangle}=\langle\turmac{x}{z},w\rangle$; moroever,
$\timetm{e_\mathit{tens}}{x}\leq q$ and 
$\timetm{y}{\langle z,w\rangle}\leq\timetm{x}{z}+r$.
Then, take $\psi\in |M|$ such that
$\normpar{M}{\psi}\geq p+r$, put 
$\sigma=\psi+\varphi+\mu$, where $\normpar{M}{\mu}\geq\cp$. Suppose
$\maj{A\otimes C}{\alpha}{\langle z,w\rangle}{(a,c)}$.
By definition, there are $\beta,\gamma$ such that
$$
\begin{array}{c}
\maj{A}{\beta}{z}{a}\\
\maj{C}{\gamma}{w}{c}\\
\alpha\geq_M\beta+\gamma 
\end{array}
$$
By hypothesis, there are $\delta,t$ such that
$$
\begin{array}{c}
\maj{B}{\delta}{t}{f(a)}\\
\delta\leq_M\varphi+\beta\\
\turmac{e}{z}=t\\
\timetm{e}{z}\leq \distpar{M}{\delta}{\varphi+\beta}
\end{array}
$$
Then,
$\maj{B\otimes C}{\gamma+\delta+\mu}{\langle t,w\rangle}{(f(a),c)}$.
Moreover,
$$
\gamma+\delta+\mu\leq_M \gamma+\varphi+\beta+\mu\leq_M \alpha+\varphi+\mu\leq_M \alpha+\sigma
$$
Finally:
\begin{eqnarray*}
\timetm{y}{\langle z,w\rangle}&\leq&\timetm{x}{z}+r\\
 &\leq& \distpar{M}{\delta}{\varphi+\beta}+\normpar{M}{\psi}\\
 &\leq& \distpar{M}{\delta}{\varphi+\beta+\psi}\\
 &\leq& \distpar{M}{\gamma+\delta+\mu}{\gamma+\varphi+\beta+\mu+\psi}\\
 &=& \distpar{M}{\gamma+\delta+\mu}{\gamma+\beta+\sigma}\\
 &=& \distpar{M}{\gamma+\delta+\mu}{\alpha+\sigma}
\end{eqnarray*}
This concludes the proof, since $tens:\morphism{(A\linear B)}{(f,g)}{\xi}{(A\otimes C)\linear(B\otimes C)}$
where $\xi$ is such that $\normpar{M}{\xi}\geq q+|e_\mathit{tens}|$.\hfill $\Box$
.\hfill $\Box$
\end{proof}
Thus:}{}
\begin{lemma}
Length spaces and their morphisms form a symmetric monoidal closed
category with tensor and linear implication given as above.
\end{lemma}
A length space $I$ is defined by $|I|=\{0\}$ and
$\maj{A}{\alpha}{e}{0}$ when $\normpar{M}{\alpha}\geq |e|$. For each
length space $A$ there are isomorphisms $A\otimes I\simeq A$ and a
unique morphism $A\rightarrow I$. The latter serves to justify full 
weakening.\par
\condinc{
For every resource monoid $M$, there is a length space 
$B_M=(\{0,1\}^*,\vdashp{B_M})$ where 
$\maj{B_M}{\alpha}{\Phi(t)}{t}$ whenever $\normpar{M}{\alpha}\geq |t|$.
The function $s_0$ (respectively, $s_1$) from $\{0,1\}^*$ to itself 
which appends $0$ (respectively, $1$) to the left
of its argument can be computed in constant time on the abstract
computational model and, as a consequence, is a morphism from $B_M$ 
to itself.}
{For every resource monoid $M$, there is a length space 
$B_M=(\{0,1\}^*,\vdashp{B_M})$ where 
$\maj{B_M}{\alpha}{\overline{t}}{t}$ whenever $\overline{t}$ is
a realizer for $t$ and $\normpar{M}{\alpha}\geq |\overline{t}|$.
The function $s_0$ (respectively, $s_1$) from $\{0,1\}^*$ to itself 
which appends $0$ (respectively, $1$) to the left
of its argument can be computed in constant time in our
computational model and, as a consequence, is a morphism from $B_M$ 
to itself.}
\subsection{Interpreting Multiplicative Affine Logic}\label{bloed}
We can now formally show that second order multiplicative
affine logic (i.e. multiplicative linear logic plus
full weakening) can be interpreted inside the category of
length spaces on any monoid $M$. Doing this will simplify
the analysis of richer systems presented in following sections.
Formulae of (intuitionistic) multiplicative affine logic 
are generated by the following productions:
$$
A::=\alpha\;|\;A\linear A\;|\;A\otimes A\;|\;\forall\alpha.A
$$
where $\alpha$ ranges over a countable set of atoms.
Rules are reported in figure~\ref{figure:MAL}.
\begin{figure*}
\begin{center}
\fbox{
\begin{minipage}{.98\textwidth}
{\bf Identity, Cut and Weakening}.
$$
\begin{array}{lcr}
\infer[I]{A\vdash A}{} & \;\; 
\infer[U]{\Gamma,\Delta\vdash B}{\Gamma\vdash A & \Delta,A\vdash B} &\;\;
\infer[W]{\Gamma,B\vdash A}{\Gamma\vdash A} 
\end{array}
$$
{\bf Multiplicative Logical Rules}.
$$
\begin{array}{ccccccc}
\infer[L_\otimes]{\Gamma,A\otimes 
   B\vdash C}{\Gamma,A,B\vdash C} &\,&
\infer[R_\otimes]{\Gamma,\Delta\vdash A\otimes B}{\Gamma\vdash A & \Delta\vdash B}  &\,&
\infer[L_\multimap]{\Gamma,\Delta,A\multimap 
   B\vdash C}{\Gamma\vdash A & \Delta,B\vdash C} &\,&
\infer[R_\multimap]{\Gamma\vdash A\multimap B}{\Gamma,A\vdash B}  
\end{array}
$$
{\bf Second Order Logical Rules}.
$$
\begin{array}{lcr}
\infer[L^\forall]{\Gamma,\forall\alpha.A\vdash B}{\vdash \Gamma,A[C/\alpha]\vdash B} & \;\; &
\infer[R^\forall]{\Gamma\vdash \forall\alpha.A}{\Gamma\vdash A & \alpha\notin\mathit{FV}(\Gamma)} 
\end{array}
$$
\end{minipage}}
\caption{Intuitionistic Multiplicative Affine Logic}\label{figure:MAL}
\end{center}
\end{figure*}
A \emph{realizability environment} is a partial function assigning length spaces (on the
same resource monoid) to atoms. 
Realizability semantics $\reasem{A}{\eta}$
of a formula $A$ on the realizability environment $\eta$ is defined by induction
on $A$:
\begin{eqnarray*}
\reasem{\alpha}{\eta} &=&\eta(\alpha)\\
\reasem{A\otimes B}{\eta} &=& 
  \reasem{A}{\eta}\otimes\reasem{B}{\eta}\\
\reasem{A\linear B}{\eta} &=& 
  \reasem{A}{\eta}\linear\reasem{B}{\eta}\\
\reasem{\forall\alpha.A}{\eta} &=& (|\reasem{\forall\alpha.A}{\eta}|,
\vdashp{\reasem{\forall\alpha.A}{\eta}})
\end{eqnarray*}
where 
\begin{eqnarray*}
|\reasem{\forall\alpha.A}{\eta}|&=&\prod_{C\in\mathscr{U}}|\reasem{A}{\eta[\alpha\rightarrow C]}|\\
\maj{\reasem{\forall\alpha.A}{\eta}}{\alpha}{e}{a}&\Longleftrightarrow&\forall C.
\maj{\reasem{A}{\eta[\alpha\rightarrow C]}}{\alpha}{e}{a}
\end{eqnarray*}
Here $\mathscr{U}$ stands for the class of all length spaces. A little
care is needed when defining the product since strictly speaking it
does not exist for size reasons. The standard way out is to let the
product range over those length spaces whose underlying set equals the
set of equivalence classes of a partial equivalence relation on $L$. As
already mentioned, every length space is isomorphic to one such. When
working with the product one has to insert these isomorphisms in
appropriate places which, however, we elide to increase readability.\par 
If $n\geq 0$ and $A_1,\ldots, A_n$ are formulas,
the expression $\reasem{A_1\otimes\ldots\otimes A_n}{\eta}$ stands
for $I$ if $n=0$ and 
$\reasem{A_1\otimes\ldots\otimes A_{n-1}}{\eta}\otimes\reasem{A_n}{\eta}$
if $n\geq 1$. 
\section{Elementary Length Spaces}\label{sect:els}
In this section, we define a resource monoid $\mathcal{L}$
such that elementary affine logic can be interpreted in the
category of length spaces on $\mathcal{L}$. We then (re)prove
that functions representable in \EAL\ are elementary time
computable.\par
A \emph{list} is either $\emptylist$ or
$\cons{n}{l}$ where $n\in\N$ and
$l$ is itself a list. 
The sum $l+h$ of two lists $l$ and $h$ is defined as
follows, by induction on $l$:
\begin{eqnarray*} 
\emptylist + h = h + \emptylist &=& h\\
\cons{n}{l}+\cons{m}{h}&=&\cons{n+m}{l+h}
\end{eqnarray*}
For every $e\in\N$, binary relations $\leq_e$ on lists can be defined as follows
\begin{varitemize}
  \item
  $\emptylist\leq_e l$;
  \item
  $\cons{n}{l}\leq_e\cons{m}{h}$ iff there is $d\in\N$ such that
  \begin{numlist}
     \item
     $n\leq 3^e(m+e)-d$;
     \item
     $l\leq_d h$.
  \end{numlist}
\end{varitemize}
For every $e$ and for every lists $l$ and $h$ with $l\leq_e h$, we define 
the natural number $\distpar{e}{l}{h}$ as follows:
\condinc{
\begin{eqnarray*}
\distpar{e}{\emptylist}{\emptylist}&=&0;\\
\distpar{e}{\emptylist}{\cons{n}{l}}&=&3^e(n+e)+\distpar{3^e(n+e)}{\emptylist}{l};\\
\distpar{e}{\cons{n}{l}}{\cons{m}{h}}&=&3^e(m+e)-n+\distpar{3^e(m+e)-n}{l}{h};
\end{eqnarray*}}{ 
$$
\begin{array}{l}
\distpar{e}{\emptylist}{\emptylist}=0;\\
\distpar{e}{\emptylist}{\cons{n}{l}}=
3^e(n+e)+\distpar{3^e(n+e)}{\emptylist}{l};\\
\distpar{e}{\cons{n}{l}}{\cons{m}{h}}=
3^e(m+e)-n+\distpar{3^e(m+e)-n}{l}{h};
\end{array}
$$}
Given a list $l$, $!l$ stands for the list $\cons{0}{l}$. The depth
$\depth{l}$ of a list $l$ is defined by induction on $l$:
$\depth{\emptylist}=0$ while
$\depth{\cons{n}{l}}=\depth{l}+1$.
$|l|$ stands for the maximum integer appearing inside $l$, i.e.
$|\emptylist|=0$ and
$|\cons{n}{l}|=\max\{n,|l|\}$. 
For every natural number $n$, $\basiclist{n}$ stands for
$\cons{n}{\emptylist}$.\par
\condinc{We can now verify that all the necessary conditions required by the
definition of a resource monoid are satisfied. To do this, we need a 
number of preliminary results, which can all be proved by simple
inductions and case-analysis:
\begin{lemma}[Compatibility]\label{lemma:elscompat}
$\emptylist\leq_e l$ for every $l$. Moreover,
if $l,h,j$ are lists and $l\leq_e h$, then
$l+j\leq_e h+j$.
\end{lemma}  
\begin{proof}
The first claim is trivial. To prove the second,
we proceed by an induction on $j$. If $j=\emptylist$,
then $l+j=l\leq_e h=h+j$. Now, suppose $j=\cons{n}{g}$.
If $h=\emptylist$, then
$l=\emptylist$ and, clearly $l+j=j\leq_e j=h+j$.
If $l=\emptylist$, we have to prove that $j\leq_e h+j$. 
Let $h=\cons{m}{f}$; then
\begin{eqnarray*}
n&\leq& n+m\leq 3^e(n+m+e)-0\\
g&\leq_0& g+f
\end{eqnarray*}
which means $j\leq_e h+j$.
Finally, suppose $l=\cons{m}{f}$, $h=\cons{p}{r}$. Then we know that
\begin{eqnarray*}
m&\leq& 3^e(p+e)-d\\
f&\leq_d& r
\end{eqnarray*}
But then, by inductive hypothesis,
\begin{eqnarray*}
m+n&\leq& 3^e(p+e)+n-d\leq 3^e(p+n+e)-d\\
f+g&\leq_d& r+g
\end{eqnarray*}
which yields $l+j\leq_e h+j$.\qed
\end{proof}
\begin{lemma}[Transitivity]\label{lemma:elstrans}
If $l,h,j$ are lists and $l\leq_e h$, $h\leq_d j$, then
$l\leq_{d+e} j$.
\end{lemma}
\begin{proof}
We can suppose all the involved lists to be different
from $\emptylist$, since all the other cases are trivial.
$l=\cons{n}{g}$, $h=\cons{m}{f}$ and $j=\cons{p}{r}$.
From the hypothesis, we have 
\begin{eqnarray*}
n&\leq& 3^e(m+e)-c\\
m&\leq& 3^d(p+d)-b\\
g&\leq_c& f\\
f&\leq_b& r
\end{eqnarray*}
But then, by inductive hypothesis, we get
\begin{eqnarray*}
n&\leq& 3^e(m+e)-c\leq 3^e(3^d(p+d)-b+e)-c\leq 3^e3^d(p+d+e)-b-c=3^{e+d}(p+d+e)-(b+c)\\\
g&\leq_{c+b}& r\\
\end{eqnarray*}
This means $l\leq_{d+e} j$.\qed
\end{proof}
\begin{lemma}\label{lemma:elsdistcompat}
if $l,h,j$ are lists and $l\leq_e h$, then
$\distpar{e}{l}{h}\leq\distpar{e}{l+j}{h+j}$
\end{lemma}
\begin{proof}
We proceed by an induction on $j$. If $j=\emptylist$,
then $l+j=l$ and $h+j=h$. Now, suppose $j=\cons{n}{g}$.
If $h=\emptylist$, then
$l=\emptylist$ and, clearly $l+j=j=h+j$.
If $l=\emptylist$, let $h=\cons{m}{f}$; then
\begin{eqnarray*}
\distpar{e}{l}{h}&=&\distpar{e}{\emptylist}{h}=3^e(m+e)+\distpar{3^e(m+e)}{\emptylist}{f}\\
  &\leq&3^e(m+e)+3^en-3^en+\distpar{3^e(m+e)+3^en-3^en}{g}{g+f}\\
  &\leq&3^e(m+n+e)-n+\distpar{3^e(m+n+e)-n}{g}{g+f}\\
  &=&\distpar{e}{j}{h+j}=\distpar{e}{l+j}{h+j}
\end{eqnarray*}
Finally, suppose $l=\cons{m}{f}$, $h=\cons{p}{r}$. Then we know that
\begin{eqnarray*}
\distpar{e}{l}{h}&=&3^e(m+e)-p+\distpar{3^e(m+e)-p}{f}{r}\\
  &\leq&3^e(m+e)-p+\distpar{3^e(m+e)-p}{f+g}{r+g}\\
  &\leq&3^e(m+e)+3^en-(n+p)+\distpar{3^e(m+e)+3^en-n-p}{f+g}{r+g}\\
  &=&3^e(m+n+e)-(n+p)+\distpar{3^e(m+n+e)-(n+p)}{f+g}{r+g}\\
  &=&\distpar{e}{l+j}{h+j}
\end{eqnarray*}
\end{proof}
\begin{lemma}\label{lemma:elsdisttrans}
If $l,h,j$ are lists and $l\leq_e h$, $h\leq_d j$, then
$\distpar{e}{l}{h}+\distpar{d}{h}{j}\leq\distpar{e+d}{l}{j}$.
\end{lemma}
\begin{proof}
If either $h=\emptylist$ or $j=\emptylist$, then the thesis
is trivial. So suppose $h=\cons{n}{g}$ and $j=\cons{m}{f}$.
If $l=\emptylist$, then
\begin{eqnarray*}
\distpar{e}{l}{h}+\distpar{d}{h}{j}&=&3^e(n+e)+\distpar{3^e(n+e)}{\emptylist}{g}
  +3^d(m+d)-n+\distpar{3^e(m+d)-n}{g}{f}\\
&\leq&3^e(n+e)+3^d(m+d)-n+\distpar{3^e(n+e)+3^d(m+d)-n}{\emptylist}{f}\\
&\leq&(3^e-1)n+3^ee+3^d(m+d)+\distpar{(3^e-1)n++3^ee+3^d(m+d)}{\emptylist}{f}\\
&\leq&(3^e-1)3^d(m+d)+3^ee+3^d(m+d)+\distpar{(3^e-1)3^d(m+d)+3^ee+3^d(m+d)}{\emptylist}{f}\\
&=&3^{d+e}(m+d+e)+\distpar{3^{d+e}(m+d+e)}{\emptylist}{f}\\
&=&\distpar{e+d}{l}{j}
\end{eqnarray*}
If $l=\cons{p}{r}$, then
\begin{eqnarray*}
\distpar{e}{l}{h}+\distpar{d}{h}{j}&=&3^e(n+e)-p+\distpar{3^e(n+e)-p}{r}{g}
  +3^d(m+d)-n+\distpar{3^d(m+d)-n}{g}{f}\\
&\leq&3^e(n+e)-p+3^d(m+d)-n+\distpar{3^e(n+e)-p+3^d(m+d)-n}{r}{f}\\
&\leq&(3^e-1)n+3^ee+3^d(m+d)-p+\distpar{(3^e-1)n+3^ee+3^d(m+d)-p}{r}{f}\\
&\leq&(3^e-1)3^d(m+d)+3^ee+3^d(m+d)-p+\distpar{(3^e-1)3^d(m+d)+3^ee+3^d(m+d)-p}{r}{f}\\
&=&3^{d+e}(m+d+e)-p+\distpar{3^{d+e}(m+d+e)-p}{r}{f}\\
&=&\distpar{e+d}{l}{j}
\end{eqnarray*}
This concludes the proof.\qed
\end{proof}}{Relation $\leq_0$ and function $\dist_0$ can
be used to build a resource monoid on lists.}
$|\lists|$ will denote the set of all lists, while $\leq_\lists,\dist_\lists$ will 
denote $\leq_0$ and $\dist_0$, respectively.
\begin{lemma}\label{lemma:elsresource}
$\lists=(|\lists|,+,\leq_\lists,\dist_\lists)$ is a resource monoid.
\end{lemma}
\condinc{\begin{proof}
$(\lists,+)$ is certainly a monoid. Compatibility of $\leq_\lists$ follows from
lemmas~\ref{lemma:elscompat} and~\ref{lemma:elstrans}. The two required
property on $\dist_\lists$ come directly from lemmas~\ref{lemma:elsdistcompat} 
and~\ref{lemma:elsdisttrans}. If $n\in\N$, observe that $\normpar{\lists}{\cons{n}{\emptylist}}=n$. 
This concludes the proof.\qed
\end{proof}}{}
An \emph{elementary length space} is a length space on the resource
monoid $(|\lists|,+,\leq_\lists,\dist_\lists)$.
Given an elementary length space $A=(|A|,\vdashp{A})$, we can build the
length space $!A=(|A|,\vdashp{!A})$, where $\maj{!A}{l}{e}{a}$
iff $\maj{A}{h}{e}{a}$ and $l\geq_\lists !h$. The construction
$!$ on elementary length spaces serves to capture the exponential modality
of elementary affine logic. Indeed, the following two results prove
the existence of morphisms and morphisms-forming rules
precisely corresponding to axioms and rules from \EAL.
\condinc{
\begin{lemma}\label{lemma:ealcontr}
For every $e\in\N$ and for every $l\in\lists$, 
$l+l\leq_1 l$ and $\distpar{e+1}{l+l}{l}\geq
\distpar{e}{0}{l}$.
\end{lemma}
\begin{proof} 
The inequality $l+l\leq_1 l$ can be proved
by induction on $l$. The base case is trivial.
If $l=\cons{n}{h}$, then
\begin{eqnarray*}
  n+n&\leq&3n+3-1=3^1(n+1)-1\\
  h+h\leq_1&h
\end{eqnarray*}
The second inequality can be proved
by induction on $l$, too. The base case is trivial.
If $l=\cons{n}{h}$, observe that
\begin{eqnarray*}
\distpar{e+1}{l+l}{l}&=&3^{e+1}(n+e+1)-2n+\distpar{3^{e+1}(n+e+1)-2n}{h+h}{h}\\
\distpar{e}{0}{l}&=&3^{e}(n+e)+\distpar{3^{e}(n+e)}{0}{h}
\end{eqnarray*}
But
\begin{eqnarray*}
3^{e+1}(n+e+1)-2n&=& 3^e(n+e+1)+2(3^e)(n+e+1)-2n\\
&\geq& 3^e(n+e+1)+2n-2n\geq 3^e(n+e)+1
\end{eqnarray*}
This concludes the proof.
\qed
\end{proof}}{}
\condinc{
\begin{lemma}[Basic Maps]
Given elementary length spaces $A,B$, there are morphisms:
\begin{eqnarray*}
\mathit{contr}&:&!A\rightarrow!A\otimes!A\\
\mathit{distr}&:&!A\otimes!B\rightarrow !(A\otimes B)
\end{eqnarray*}
where
$\mathit{contr}(a)=(a,a)$ and 
$\mathit{distr}(a,b)=(a,b)$
\end{lemma}}{
\begin{lemma}[Basic Maps]
Given elementary length spaces $A,B$, there are morphisms
$\mathit{contr}:!A\rightarrow!A\otimes!A$ and
$\mathit{distr}:!A\otimes!B\rightarrow !(A\otimes B)$
where
$\mathit{contr}(a)=(a,a)$ and
$\mathit{distr}(a,b)=(a,b)$.
\end{lemma}}
\condinc{
\begin{proof}
We know $\{e_\mathit{contr}\}(d)$ takes time $|d|+p$,
where $p$ is a constant. Then, let $l,h\in \lists$ be
such that $\normpar{\lists}{l}\geq p+|e_\mathit{contr}|$,
$\normpar{\lists}{h}\geq\cp$. Define $l_\mathit{contr}$
to be $l+h+\basiclist{1}$. Clearly, 
$\normpar{\lists}{l_\mathit{contr}}\geq |e_\mathit{contr}|$
Now, let $\maj{!A}{j}{d}{a}$. This implies that
$j\geq_{\lists}!k$ where $\maj{A}{k}{d}{a}$.
Then:
\begin{eqnarray*}
h+!k+!k&\geq_{\lists}& !k+!k\\
\normpar{\lists}{h+!k+!k}&\geq&\normpar{\lists}{h}+\normpar{\lists}{!k}+\normpar{\lists}{!k}\\
&\geq&\cp+\normpar{\lists}{!k}+\normpar{\lists}{!k}
\end{eqnarray*}
This yields $\maj{!A\otimes !A}{h+!k+!k}{e}{(a,a)}$.
By lemma~\ref{lemma:ealcontr}, $h+!k+!k\leq_{\lists}h+!k+\basiclist{1}
\leq_{\lists}h+j+\basiclist{1}\leq_{\lists}j+l_\mathit{contr}$.
Finally,
\begin{eqnarray*}
\timetm{e_\mathit{contr}}{d}&\leq&|d|+p\leq \normpar{\lists}{k}+p
  \leq\distpar{\lists}{!k+!k}{!k+\basiclist{1}}+\normpar{\lists}{l}\\
&\leq& \distpar{\lists}{!k+!k}{!k+\basiclist{1}+l}\\
&\leq& \distpar{\lists}{!k+!k+h}{!k+\basiclist{1}+l+h}\\
&=& \distpar{\lists}{!k+!k+h}{!k+l_\mathit{contr}}
\end{eqnarray*}
This proves $\mathit{contr}$ to be a morphism.\par
Let $e_\mathit{distr}=e_\mathit{id}$. We know
$\{e_\mathit{id}\}(d)$ takes constant time,
say $p$. Then, let $l,h\in \lists$ be
such that $\normpar{\lists}{l}\geq p+|e_\mathit{distr}|$,
$\normpar{\lists}{h}\geq\cp$. $l_\mathit{distr}$ is then
defined as $l+!h$. 
Now, let $\maj{!A\otimes !B}{j}{\langle d,c\rangle}{(a,b)}$. 
This means that $j\geq !k+!i$, where 
$\maj{A}{k}{d}{a}$ and $\maj{B}{i}{c}{b}$. This in turn
means that $\maj{A\otimes B}{k+i+h}{\langle d,c\rangle}{(a,b)}$
and  $\maj{!(A\otimes B)}{!(k+i+h)}{\langle d,c\rangle}{(a,b)}$.
Moreover
$$
!(k+i+h)=!k+!i+!h\leq_\lists j+!h\leq_\lists j+l_\mathit{distr}
$$
Finally:
\begin{eqnarray*}
\timetm{e_\mathit{distr}}{\langle d,c\rangle}&\leq&p\leq\normpar{\lists}{l}\\
&\leq& \distpar{\lists}{!(k+i+h)}{j+!h}+\normpar{\lists}{l}\\
&\leq& \distpar{\lists}{!(k+i+h)}{j+!h+l}\\
&\leq& \distpar{\lists}{!(k+i+h)}{j+l_\mathit{distr}}
\end{eqnarray*}
This proves $\mathit{distr}$ to be a morphism.\qed
\end{proof}}{}
\begin{lemma}[Functoriality]
If $f:\morphism{A}{e}{\varphi}{B}$, then there is $\psi$ such
that $f:\morphism{!A}{e}{\psi}{!B}$
\end{lemma}
\condinc{\begin{proof}
Let $\theta$ be $!\varphi$ and suppose $\maj{!A}{d}{l}{a}$. Then $l\geq !h$,
where $\maj{A}{d}{h}{a}$. Observe that there must be $j,c$ such that
$\maj{B}{c}{j}{f(a)}$, $j\leq_{\lists} h+\varphi$ and
$\timetm{e}{d}\leq\distpar{\lists}{j}{h+\varphi}$.
But then $\maj{!B}{c}{!j}{f(a)}$ and, moreover
\begin{eqnarray*}
!j&\leq_{\lists}&!(h+\varphi)=!h+!\varphi\leq_{\lists}!h+\theta\\
\timetm{e}{d}&\leq&\distpar{\lists}{j}{h+\varphi}
  \leq\distpar{\lists}{!j}{!(h+\varphi)}\\
  &\leq&\distpar{\lists}{!j}{!h+!\varphi)}
  \leq\distpar{\lists}{!j}{l+\theta}
\end{eqnarray*}
This means that $f:\morphism{!A}{e}{\theta}{!B}$.\qed
\end{proof}}{}
Elementary bounds can be given on $\normpar{\lists}{l}$ depending
on $|l|$ and $\depth{l}$:
\begin{proposition}\label{prop:EALbound}
For every $n\in\N$ there is an elementary function $p_n:\N\rightarrow\N$ such
that $\normpar{\lists}{l}\leq p_{\depth{l}}(|l|)$.
\end{proposition}
\condinc{\begin{proof}
We prove a stronger statement by induction on $n$: for every $n\in\N$
there is an elementary function $q_n:\N^2\rightarrow\N$ such that for every $l,e$,
$\distpar{e}{\emptylist}{l}\leq q_{\depth{l}}(|l|,e)$. First of all, we
know that $\distpar{e}{\emptylist}{\emptylist}=0$, so $q_0$ is just the
function which always returns $0$. $q_{n+1}$ is defined from $q_n$ as
follows: $q_{n+1}(x,y)=3^y(x+y)+q_n(x,3^y(x+y))$.
Indeed:
\begin{eqnarray*}
\distpar{e}{\emptylist}{\cons{n}{l}}&=&
     3^e(n+e)+\distpar{3^e(n+e)}{\emptylist}{l}\\
  &\leq&3^e(|\cons{n}{l}|+e)+q_{\depth{l}}(|l|,3^e(n+e))\\
  &\leq&3^e(|\cons{n}{l}|+e)+q_{\depth{l}}(|\cons{n}{l}|,3^e|\cons{n}{l}|+e)\\
  &=&q_{\depth{\cons{n}{l}}}(|\cons{n}{l}|,e)
\end{eqnarray*}
At this point we just put $p_n(x)=q_n(x,0)$.\qed
\end{proof}}{}
We emphasize that Proposition~\ref{prop:EALbound} does not assert that the mapping
$(n,m)\mapsto p_n(m)$ is elementary. This, indeed, cannot be true
because we know \EAL\ to be complete for the class of elementary
functions. If, however, $A\subseteq\lists$ is such that $l\in A$
implies $\depth{l}\leq c$ for a fixed $c$, then $(l\in A)\mapsto
p_{\depth{l}}(|l|)$ is elementary and it is in this way that we will
use the above proposition.

\subsection{Interpreting Elementary Affine Logic}
\EAL\ can be obtained by endowing multiplicative affine logic
with a restricted modality. The grammar of formulae is enriched with 
a new production $A::= !A$
while modal rules are reported in figure~\ref{figure:EAL}. 
\begin{figure*}
\begin{center}
\fbox{
\begin{minipage}{.8\textwidth}
{\bf Exponential Rules and Contraction}.
$$
\begin{array}{lcr}
\infer[P]{!\Gamma\vdash !A}{\Gamma\vdash A} & \;\; & 
\infer[C]{\Gamma,!A\vdash B}{\Gamma,!A,!A\vdash B}
\end{array}
$$
\end{minipage}}
\caption{Intuitionistic Elementary Affine Logic}\label{figure:EAL}
\end{center}
\end{figure*}
Realizability semantics is extended by 
$\reasem{!A}{\eta}=!\reasem{A}{\eta}$.
\begin{theorem}\label{theo:EAL}
Elementary length spaces form a model of \EAL.
\end{theorem}
Now, consider the formula
$$
\Listeal\equiv\forall\alpha.!(\alpha\linear\alpha)\linear!(\alpha\linear\alpha)
\linear!(\alpha\linear\alpha)
$$
Binary lists can be represented as cut-free proofs
with conclusion $\Listeal$. Suppose you have a proof 
$\pi:!^j\Listeal\linear !^k\Listeal$. 
From the denotation $\reasem{\pi}{}$ we
can build a morphism $g$ from $\reasem{\Listeal}{}$ to $B_\lists$ by internal
application to $\varepsilon,s_0,s_1$. This map then induces a function 
$f: B\rightarrow B$ as
follows: given $w\in B$, first compute a realizer for
the closed proof corresponding to it, then apply $g$ to the result. 
\begin{remark}\label{remark:bounddepth}
Notice that elements of $B_\lists$ can all be 
majorized by lists with unit depth. Similarly, elements
of $\reasem{\Listeal}{}$ corresponding to binary lists
can be majorized by lists with bounded depth. This observation
is essential to prove the following result.
\end{remark}
\begin{corollary}[Soundness]
Let $\pi$ be an \EAL\ proof with conclusion $\vdash!^j\Listeal\linear!^k\Listeal$
and let $f:L\rightarrow L$ be the function induced by $\reasem{\pi}{}$.
Then $f$ is computable in elementary time.
\end{corollary}
\condinc{
}{}
The function $f$ in the previous result 
equals the function denoted by the proof $\pi$ in the sense of
\cite{hofmann04tcs}. This intuitively obvious fact can be proved
straightforwardly but somewhat tediously using a logical relation or
similar, see also \cite{hofmann04tcs}.
\condinc{
\section{Soft Length Spaces}\label{sect:sls}
The grammar of formulae for \SAL\ is the same as the one of Elementary Affine Logic.
Rules are reported in figure~\ref{figure:SAL}. 
\begin{figure*}
\begin{center}
\fbox{
\begin{minipage}{.8\textwidth}
{\bf Exponential Rules and Contraction}.
$$
\begin{array}{lcr}
\infer[P]{!\Gamma\vdash !A}{\Gamma\vdash A} & \;\; & 
\infer[C]{\Gamma,!A\vdash B}{\Gamma,A,\ldots,A\vdash B}
\end{array}
$$
\end{minipage}}
\caption{Intuitionistic Soft Affine Logic}\label{figure:SAL}
\end{center}
\end{figure*}
We here use a resource monoid whose
underlying carrier set is $|\listsint|=|\lists|\times\N$.
The sum $(l,n)+(h,m)$ of two elements in $|\listsint|$ is defined as
$(l+h,\max\{n,m\})$. For every $e\in\N$, binary relations 
$\leq_e$ on $|\listsint|$ can be defined as follows
\begin{varitemize}
  \item
  $(\emptylist,n)\leq_0 (\emptylist,m)$ iff $n\leq m$;
  \item
  $(\emptylist,n)\leq_e (\cons{m}{l},p)$ iff there is $d\in\N$ such that
  \begin{numlist}
     \item
     $e\leq m+pd$
     \item
     $(\emptylist,n)\leq_d (l,p)$
  \end{numlist}
  \item
  $(\cons{n}{l},m)\leq_e(\cons{p}{h},q)$ iff there is $d\in\N$ such that
  \begin{numlist}
     \item
     $e+n\leq p+qd$;
     \item
     $(l,m)\leq_d (h,q)$.
  \end{numlist}
\end{varitemize}
If $\alpha=(l,n)\in|\listsint|$, then $!\alpha$ will
be the couple $(\cons{0}{l},n)\in|\listsint|$. 
If there is $e$ such that $\alpha\leq_e \beta$, then we 
will simply write $\alpha\leq_\listsint \beta$.
For every $\alpha$ and $\beta$ with $\alpha\leq_\listsint \beta$, we define 
the natural number $\distpar{\listsint}{\alpha}{\beta}$ as follows: 
\condinc{
\begin{eqnarray*}
\distpar{\listsint}{(\emptylist,n)}{(\emptylist,m)}&=&0\\
\distpar{\listsint}{(\emptylist,n)}{(\cons{m}{l},p)}&=&m+p\distpar{\listsint}{(\emptylist,n)}{(l,p)}\\
\distpar{\listsint}{(\cons{n}{l},m)}{(\cons{p}{h},q)}&=&p-n+q\distpar{\listsint}{(l,m)}{(h,q)}
\end{eqnarray*}
}{
$$
\begin{array}{l}
\distpar{\listsint}{(\emptylist,n)}{(\emptylist,m)}=0;\\
\distpar{\listsint}{(\emptylist,n)}{(\cons{m}{l},p)}=
m+p\distpar{\listsint}{(\emptylist,n)}{(l,p)};\\
\distpar{\listsint}{(\cons{n}{l},m)}{(\cons{p}{h},q)}=
p-n+q\distpar{\listsint}{(l,m)}{(h,q)};
\end{array}
$$
}
Analogously, we can define $\distpar{\listsint}{\alpha}{\beta}$ simply as the maximum integer $e$
such that $\alpha\leq_e \beta$. $|\alpha|$ is the maximum integer appearing inside $\alpha$, i.e. 
$|(l,n)|=\max\{|l|,m\}$. The depth $\depth{\alpha}$ of $\alpha=(l,n)$ is $\depth{l}$.
\condinc{
\begin{lemma}[Compatibility]\label{lemma:slscompat}
$(\emptylist,0)\leq_0 \alpha$ for every $\alpha$. Moreover,
if $\alpha,\beta,\gamma\in|\listsint|$ and $\alpha\leq_e \beta$, then
$\alpha+\gamma\leq_e \beta+\gamma$. 
\end{lemma}   
\begin{proof} 
The first claim is trivial. To prove the second,
we proceed by an induction on the structure of the
first component of $\gamma$. We just consider
the case where the first components of $\alpha,\beta,\gamma$ are
all different from $\emptylist$. So, suppose
$\alpha=(\cons{n}{l},m)$, $\beta=(\cons{p}{h},q)$, $\gamma=(\cons{r}{j},s)$.
By hypothesis, we get $d\in\N$ such that
\begin{eqnarray*}
e+n&\leq&p+dq\\
(l,m)&\leq_d&(h,q)
\end{eqnarray*}
Then, $e+n+r\leq p+r+dq\leq p+r+d\max\{q,s\}$
and, by induction hypothesis, $(l+j,\max\{m,s\})\leq_d(h+j,\max\{q,s\})$.
This implies that $\alpha+\gamma\leq_e \beta+\gamma$.\qed
\end{proof}
\begin{lemma}[Transitivity]\label{lemma:slstrans}
If $\alpha,\beta,\gamma\in|\listsint|$ are lists and $\alpha\leq_e \beta$, $\beta\leq_d \gamma$, then
$\alpha\leq_{d+e} \gamma$.
\end{lemma}
\begin{proof}
We go by induction on the structure of the first component
of $\gamma$ and we suppose the first components of $\alpha,\beta,\gamma$ to be different
from $\emptylist$. So, let
$\alpha=(\cons{n}{l},m)$, $\beta=(\cons{p}{h},q)$ and $\gamma=(\cons{r}{j},s)$.
From the hypothesis, there are $c,b\in\N$ such that 
\begin{eqnarray*}
e+n&\leq&p+cq\\
d+p&\leq&r+bs\\
(l,m)&\leq_c&(h,q)\\
(h,q)&\leq_b&(j,s)
\end{eqnarray*}
But then, by inductive hypothesis, we get
\begin{eqnarray*}
(e+d)+n&\leq& d+p+cq\leq r+bs+cq\leq r+(b+c)s\\
(l,m)&\leq_{c+b}& (j,s)
\end{eqnarray*}
which yields $\alpha\leq_{d+e} \gamma$.\qed
\end{proof}
\begin{lemma}\label{lemma:slsdistcompat}
if $\alpha,\beta,\gamma\in\listsint$ and $\alpha\leq_e \beta$, then
$\distpar{\listsint}{\alpha}{\beta}\leq\distpar{\listsint}{\alpha+\gamma}{\beta+\gamma}$
\end{lemma}
\begin{proof}
This is trivial in view of~\ref{lemma:slscompat} and
the fact that $\distpar{\listsint}{\alpha}{\beta}$ is just
$\max\{e\in\N\;|\;\alpha\leq_e \beta\}$.\qed
\end{proof}
\begin{lemma}\label{lemma:slsdisttrans}
If $\alpha,\beta,\gamma\in\listsint$ and $\alpha\leq_e \beta$, $\beta\leq_d \gamma$, then
$\distpar{e}{\alpha}{\beta}+\distpar{d}{\beta}{\gamma}\leq\distpar{e+d}{\alpha}{\gamma}$.
\end{lemma}
\begin{proof}
This is trivial in view of~\ref{lemma:slstrans} and
the fact that $\distpar{\listsint}{\alpha}{\beta}$ is just
$\max\{e\in\N\;|\;\alpha\leq_e \beta\}$.\qed
\end{proof}}{}
\begin{lemma}\label{lemma:slsresource}
$(\listsint,+,\leq_\listsint,\dist_\listsint)$ is a resource monoid.
\end{lemma}
\condinc{\begin{proof}
$(|\listsint|,+)$ is certainly a commutative monoid. Compatibility of $\leq_\listsint$ follows from
lemmas~\ref{lemma:slscompat} and~\ref{lemma:slstrans}. The two required
property on $\dist_\listsint$ come directly from lemmas~\ref{lemma:slsdistcompat} 
and~\ref{lemma:slsdisttrans}. If $n\in\N$, observe that $\normpar{\listsint}{(\cons{n}{\emptylist},0)}=n$. 
This concludes the proof.\qed
\end{proof}}{}  
A \emph{soft length space} is a length space on the resource
monoid $(\listsint,+,\leq_\listsint,\dist_\listsint)$.\par
Given a soft length space $A=(|A|,\vdashp{A})$, we can build the
length space $!A=(|A|,\vdashp{!A})$, where $\maj{!A}{\alpha}{e}{a}$
iff $\maj{!A}{\beta}{e}{a}$ and $\alpha\geq_\listsint !\beta$.
We write $\basiclistint{n}{m}$ for $(\cons{n}{\emptylist},m)$.\par
\condinc{
\begin{lemma}\label{lemma:salcontr}
For every $\alpha\in\listsint$ and for every $n,m\in\N$ the following
inequality holds:
$$
n.\alpha\leq_{n\normpar{\listsint}{\alpha}+m}!\alpha+\basiclistint{m}{2n}
$$
\end{lemma}
\begin{proof}
Let $\alpha=(l,p)$. We go by induction on $l$. If $l$ is $\emptylist$, then
\begin{eqnarray*}
n.\alpha&=&(\emptylist,p)\\
!\alpha+\basiclistint{m}{2n}&=&(\cons{m}{\emptylist},\max\{p,2n\})\\
n\normpar{\listsint}{\alpha}+m&=&m\\
\emptylist&\leq_0&\emptylist
\end{eqnarray*}
This implies the thesis. Moreover, if $l=\cons{q}{h}$, then
\begin{eqnarray*}
n.\alpha&=&(n.l,p)=(\cons{nq}{n.h},p)\\
!\alpha+\basiclistint{m}{2n}&=&(\cons{m}{l},\max\{p,2n\})\\
n\normpar{\listsint}{\alpha}+m&=&n(q+p\normpar{\listsint}{l,p})+m
\end{eqnarray*}
By induction hypothesis, we get
\begin{eqnarray*}
(n.h,p)&\leq_{n\normpar{\listsint}{h,p}+q}&!(h,p)+\basiclistint{q}{2n}=(l,\max\{p,2n\})\\
(n(q+p\normpar{\listsint}{l,p})+m)+nq&=&m+2nq+np\normpar{\listsint}{l,p}\\
  &\leq&m+\max\{p,2n\}(n\normpar{\listsint}{h,p}+q)
\end{eqnarray*}
from which the desired inequality easily follows.\qed
\end{proof}
\begin{lemma}[Basic Maps]
Given soft length spaces $A,B$ and a natural number
$n\geq 1$, there are morphisms:
\begin{eqnarray*}
\mathit{contr}_n&:&!A\rightarrow \overbrace{A\otimes\ldots\otimes A}^{\mbox{$n$ times}}\\
\mathit{distr}&:&!A\otimes!B\rightarrow !(A\otimes B)
\end{eqnarray*}
where
$\mathit{contr}(a)=(\overbrace{a,\ldots,a}^{\mbox{$n$ times}})$ and 
$\mathit{distr}(a,b)=(a,b)$
\end{lemma}
\begin{proof}
We define realizers $e_\mathit{contr}^n$ for every $n\geq 1$ by
induction on $n$:
\begin{eqnarray*}
e_\mathit{contr}^1&=&e_\mathit{id}\\
e_\mathit{contr}^{n+1}&=&(e_\mathit{contr}^n)^*\circ e_\mathit{contr}
\end{eqnarray*}
Clearly, $e_\mathit{contr}^n$ is a realizer
for $\mathit{contr}_n$. Moreover, 
$\timetm{e_\mathit{contr}^n}{x}\leq n|x|+q_n$, where $q_n$ does not
depend on $x$. Now, let $\psi_n$ be such that
$\normpar{\listsint}{\psi_n}\geq \cp\cdot n$ and $\varphi_\mathit{contr}^n$ be
$\basiclistint{q_n}{2n}+\psi_n$ for every $n\geq 1$. Now,
let $\maj{!A}{\alpha}{j}{a}$. This implies $\alpha\geq_\listsint !(l,m)$,
where $\maj{A}{(l,m)}{j}{a}$. Notice that
$$
\maj{\underbrace{A\otimes\ldots\otimes A}_{\mbox{$n$ times}}}{n.(l,m)+\psi_n}{\langle \overbrace{j,\ldots,j}^{\mbox{$n$ times}}\rangle}{(\overbrace{a,\ldots,a}^{\mbox{$n$ times}})}
$$
By lemma~\ref{lemma:salcontr},
we finally get
\begin{eqnarray*}
n.(l,m)+\psi_n&\leq_\listsint& !(l,m)+\basiclistint{q_n}{2n}+\psi_n\\
&=&!(l,m)+\varphi_\mathit{contr}^n\leq \varphi_\mathit{contr}^n+\alpha\\
\timetm{e_\mathit{contr}^n}{j}&\leq&n|j|+q_n\\
&\leq&n\normpar{\listsint}{l,m}+q_n\\
&\leq&\distpar{\listsint}{n.(l,m)}{!(l,m)+\basiclistint{q_n}{2n}}\\
&\leq&\distpar{\listsint}{n.(l,m)}{(\cons{q_n}{l},\max\{m,2n\})}\\
&\leq&\distpar{\listsint}{n.(l,m)+\psi_n}{(\cons{q_n}{l},\max\{m,2n\})+\psi_n}\\
&\leq&\distpar{\listsint}{n.(l,m)+\psi_n}{\basiclistint{q_n}{2n}+\alpha+\psi_n}\\
&\leq&\distpar{\listsint}{n.(l,m)+\psi_n}{\alpha+\varphi^n_\mathit{contr}}
\end{eqnarray*}
This proves each $e_\mathit{contr}^n$ to be a morphism.\par
Let $e_\mathit{distr}=e_\mathit{id}$. We know
$\{e_\mathit{id}\}(d)$ takes constant time,
say $p$. Then, let $\psi,\mu\in \listsint$ be
such that $\normpar{\listsint}{\psi}\geq p+|e_\mathit{distr}|$,
$\normpar{\listsint}{\mu}\geq\cp$. $\varphi_\mathit{distr}$ is then
defined as $\psi+!\mu$. 
Now, let $\maj{!A\otimes !B}{\alpha}{\langle d,c\rangle}{(a,b)}$. 
This implies $\alpha\geq !\beta+!\gamma$, where 
$\maj{A}{\beta}{d}{a}$ and $\maj{B}{\gamma}{c}{b}$. This in turn
implies $\maj{A\otimes B}{\beta+\gamma+\mu}{\langle d,c\rangle}{(a,b)}$
and  $\maj{!(A\otimes B)}{!(\beta+\gamma+\mu)}{\langle d,c\rangle}{(a,b)}$. 
Moreover
$$
!(\beta+\gamma+\mu)=!\beta+!\gamma+!\mu\leq_\lists \alpha+!\mu\leq_\lists \alpha+\varphi_\mathit{distr}
$$
Finally:
\begin{eqnarray*}
\timetm{e_\mathit{distr}}{\langle d,c\rangle}&\leq&p\leq\normpar{\lists}{\psi}\\
&\leq& \distpar{\lists}{!(\beta+\gamma+\mu)}{\alpha+!\mu}+\normpar{\lists}{\psi}\\
&\leq& \distpar{\lists}{!(\beta+\gamma+\mu)}{\alpha+!\mu+\psi}\\
&\leq& \distpar{\lists}{!(\beta+\gamma+\mu)}{\alpha+\varphi_\mathit{distr}}
\end{eqnarray*}
This proves $\mathit{distr}$ to be a morphism.\qed
\end{proof}
\begin{lemma}[Functoriality]
If $f:\morphism{A}{e}{\varphi}{B}$, then there is $\psi$ such
that $f:\morphism{!A}{e}{\psi}{!B}$
\end{lemma}
\begin{proof}
Let $\theta$ be $!\varphi$ and suppose $\maj{!A}{\alpha}{d}{a}$. Then $\alpha\geq !\beta$,
where $\maj{A}{\beta}{d}{a}$. Observe that there must be $\gamma,c$ such that
$\maj{B}{\gamma}{c}{f(a)}$, $\gamma\leq_{\lists} \beta+\varphi$ and
$\timetm{e}{d}\leq\distpar{\lists}{\gamma}{\beta+\varphi}$.
But then $\maj{!B}{!\gamma}{c}{f(a)}$ and, moreover
\begin{eqnarray*}
!\gamma&\leq_{\lists}&!(\beta+\varphi)=!\beta+!\varphi\leq_{\lists}!\beta+\theta\\
\timetm{e}{d}&\leq&\distpar{\lists}{\gamma}{\beta+\varphi}
  \leq\distpar{\lists}{!\gamma}{!(\beta+\varphi)}\\
  &\leq&\distpar{\lists}{!\gamma}{!\beta+!\varphi}
  \leq\distpar{\lists}{!\gamma}{\alpha+\theta}
\end{eqnarray*}
This implies $f:\morphism{!A}{e}{\theta}{!B}$.\qed
\end{proof}
}{The following two results can be proved with techniques similar
to those from Proposition~\ref{prop:EALbound} and Theorem~\ref{theo:EAL}}
\begin{proposition}
For every $n\in\N$ there is a polynomial $p_n:\N\rightarrow\N$ such
that $\normpar{\listsint}{\alpha}\leq p_{\depth{\alpha}}(|\alpha|)$
for every $\alpha\in |\listsint|$.
\end{proposition}
\condinc{\begin{proof}
We go by induction on $n$. First of all, we
know that $\distpar{\listsint}{(\emptylist,0)}{(\emptylist,m)}=0$, so 
$p_0$ is just the function which always returns $0$. 
$p_{n+1}$ is defined from $p_n$ as follows: $p_{n+1}(x)=x+xp_n(x)$.
Indeed:
\begin{eqnarray*}
\distpar{\listsint}{(\emptylist,0)}{(\cons{n}{l},m)}&=&
     n+m\distpar{\listsint}{(\emptylist,0)}{(l,m)}\\
  &\leq&|(\cons{n}{l},m)|+|(\cons{n}{l},m)|p_{\depth{(l,m)}}(|(\cons{n}{l},m)|)\\
  &=&p_{\depth{(\cons{n}{l},m)}}((\cons{n}{l},m)).
\end{eqnarray*}
This concludes the proof.\qed
\end{proof}}{}
Again, we do not claim that $(n,m)\mapsto p_n(m)$ is a polynomial
(c.f. Remark~\ref{remark:bounddepth}).
\begin{theorem}
Soft length spaces form a model of \SAL.
\end{theorem}
Binary lists can be represented in \SAL\ as cut-free proofs
with conclusion
$$
\Listsal\equiv\forall\alpha.!(\alpha\linear\alpha)\linear !(\alpha\linear\alpha)
\linear(\alpha\linear\alpha)
$$
\begin{corollary}[Soundness]
Let $\pi$ be an \SAL\ proof with conclusion $\vdash !^j\Listsal\linear !^k\Listsal$
and let $f:L\rightarrow L$ be the function induced by $\reasem{\pi}{}$.
Then $f$ is computable in polynomial time.
\end{corollary}
\section{Light Length Spaces}\label{sect:lls}
The grammar of formulae for Light Affine Logic is the one from Elementary Affine Logic, enriched
with a new production $A::=\S A$. Rules are reported in figure~\ref{figure:LAL}. 
\begin{figure*}
\begin{center}
\fbox{
\begin{minipage}{.8\textwidth}
{\bf Exponential Rules and Contraction}.
$$
\begin{array}{lccr}
\infer[P_\S]{\S\Gamma,!\Delta\vdash \S A}{\Gamma,\Delta\vdash A} \;\; & 
\infer[P_!^1]{!A\vdash !B}{A\vdash B} \; & \; 
\infer[P_!^2]{\vdash !A}{\vdash A} & \;\;  
\infer[C]{\Gamma,!A\vdash B}{\Gamma,!A,!A\vdash B}
\end{array}
$$
\end{minipage}}
\caption{Intuitionistic Light Affine Logic}\label{figure:LAL}
\end{center}
\end{figure*}
Light length spaces are a model of Light Affine
Logic. The underlying resource monoid 
is more complex than the ones we encountered
so far. This complexity is a consequence
of the strange behaviour of 
modality $!$, which is functorial but does
not distribute over tensor (i.e. $!(A\otimes B)
\not\cong !A\otimes !B$).\par
A \emph{tree} is either $\emptytree$ or
a triple $\node{n}{t}{T}$ where $n\in\N$,
$t$ is itself a tree and $T$ is a finite
nonempty set of trees. $|\trees|$ is the
set of all trees. We write $\basictree{n}$ for
the tree $\node{n}{\emptytree}{\{\emptytree\}}$.
The sum $t+s$ of two trees $t$ and $s$ is defined as
follows, by induction on $n$:
\condinc{
\begin{eqnarray*} 
\emptytree + t &=& t + \emptytree=t;\\
\node{n}{t}{T}+\node{m}{u}{U}&=&\node{n+m}{t+u}{T\cup U};
\end{eqnarray*}}{
$$
\begin{array}{l} 
\emptytree + t = t + \emptytree=t;\\
\node{n}{t}{T}+\node{n}{u}{U}=\node{n+m}{t+u}{T\cup U};
\end{array}
$$}
Here, more sophisticated techniques are needed. 
For every $n,e\in\N$, binary relations $\leq_e^n$ on trees can be defined as follows
\begin{varitemize}
  \item
  $t\leq_e^0 u$ for every $t,u\in |\trees|$;
  \item
  $\emptytree\leq_e^{n+1} t$ for every $t\in |\trees|$;
  \item
  $\node{m}{t}{T}\leq_e^{n+1}\emptytree$ iff there is $d\in\N$ such that 
  \begin{numlist}
    \item
    $m\leq e-d$;
    \item
    $t\leq_{d^2}^n\emptytree$;
    \item
    For every $s\in T$, $s\leq_{d}^n\emptytree$.
  \end{numlist}
  \item
  $\node{m}{t}{T}\leq_e^{n+1}\node{l}{u}{U}$ iff there is $d\in\N$ such that
  \begin{numlist}
     \item
     $m\leq l+e-d$;
     \item
     There is a function $f:\{1,\ldots,d\}\rightarrow U$
     such that $t\leq_{d^2}^n u+\sum_1^d f(i)$;
     \item
     For every $s\in T$ there is $z\in U$ with $s\leq_d^n z$.
  \end{numlist}
\end{varitemize}
For every $e,n\in\N$ and for every trees $t$ and $u$ with $t\leq_e^n u$, we define 
the natural number $\distpartwo{e}{n}{t}{u}$ as follows: 
\condinc{
\begin{eqnarray*}
\distpartwo{e}{0}{t}{u}&=&0\\
\distpartwo{e}{n+1}{\emptytree}{\emptytree}&=&e+\distpartwo{e}{n}{\emptytree}{\emptytree}\\
\distpartwo{e}{n+1}{\emptytree}{\node{m}{t}{T}}&=&
  m+e+\max_f\{\distpartwo{(m+e)^2}{n}{\emptytree}{t+\sum_{i=1}^{m+e}f(i)}\}\\
\distpartwo{e}{n+1}{\node{m}{t}{T}}{\emptytree}&=&
  e-m+\distpartwo{(e-m)^2}{n}{t}{\emptytree}\\
\distpartwo{e}{n+1}{\node{m}{t}{T}}{\node{l}{u}{U}}&=&
  l+e-m+\max_f\{\distpartwo{(l+e-m)^2}{n}{t}{u+\sum_{i=1}^{l+e-m}f(i)}\}
\end{eqnarray*}}{
$$
\begin{array}{l}
\distpartwo{e}{0}{t}{u}=0;\\
\distpartwo{e}{n+1}{\emptytree}{\emptytree}=
e+\distpartwo{e}{n}{\emptytree}{\emptytree};\\
\distpartwo{e}{n+1}{\emptytree}{\node{m}{t}{T}}=m+e+\max_f\{\distpartwo{(m+e)^2}{n}{\emptytree}{t+\sum_{i=1}^{m+e}f(i)}\};\\
\distpartwo{e}{n+1}{\node{m}{t}{T}}{\emptytree}=e-m+\distpartwo{(e-m)^2}{n}{t}{\emptytree};\\
\distpartwo{e}{n+1}{\node{m}{t}{T}}{\node{l}{u}{U}}=l+e-m+\max_f\{\distpartwo{(k+e-m)^2}{n}{t}{u+\sum_{i=1}^{l+e-m}f(i)}\};
\end{array}
$$}
If $t$ is a tree, then $|t|$ is the greatest integer appearing in $t$, i.e.
$|\emptytree|=0$ and 
$|\node{n}{t}{T}|=\max\{n,|t|,\max_{u\in T}|u|\}$. 

The depth $\depth{t}$ of a tree $t$ is defined as follows:
$\depth{\emptytree}=0$ and 
\condinc{
$$\depth{\node{n}{t}{T}}=1+
\max\{\depth{t},\max_{u\in T}\depth{u}\}.$$}
{$\depth{\node{n}{t}{T}}=1+
\max\{\depth{t},\max_{u\in T}\depth{u}\}$.} 
Given a tree $t\in |\trees|$, we define $!t$ as the tree $\node{1}{\emptytree}{\{t\}}$
and $\S t$ as the tree $\node{0}{t}{\{\emptytree\}}$.
\condinc{
In this context, a notion of isomorphism between trees
is needed: we say that trees $t$ and $u$ are \emph{isomorphic} and we 
write $t\cong u$ iff for every $e,n\in\N$ and for every tree $v$
the following hold:
\begin{eqnarray*}
v\leq_e^n t&\Leftrightarrow&v\leq_e^n u\\
t\leq_e^n v&\Leftrightarrow&u\leq_e^n v\\
\distpartwo{e}{n}{v}{t}&=&\distpartwo{e}{n}{v}{u}\\
\distpartwo{e}{n}{t}{v}&=&\distpartwo{e}{n}{u}{v}
\end{eqnarray*}
\begin{lemma}
$\emptytree\cong\emptytreeone$. Moreover,
for every tree $t$, $t+\emptytree\cong t+\emptytreeone$.
\end{lemma}
\begin{proof}
We have to prove that for every $e,n\in\N$ and for every tree $v$:
\begin{eqnarray*}
v\leq_e^n \emptytree&\Leftrightarrow&v\leq_e^n \emptytreeone\\
\emptytree\leq_e^n v&\Leftrightarrow&\emptytreeone\leq_e^n v\\
\distpartwo{e}{n}{v}{\emptytree}&=&\distpartwo{e}{n}{v}{\emptytreeone}\\
\distpartwo{e}{n}{\emptytree}{v}&=&\distpartwo{e}{n}{\emptytreeone}{v}
\end{eqnarray*}
We go by induction on $n$,
considering the case where $n\geq 1$, since the base case
is trivial. First of all, observe that both
$\emptytree\leq_e^{n+1} t$ and $\emptytreeone\leq_e^{n+1} t$
for every $t$. Moreover, $\emptytree\leq_e^{n+1} \emptytree$
and $\emptytreeone\leq_e^{n+1} \emptytree$. Suppose now
that $\node{m}{t}{T}\leq_e^{n+1}\emptytree$. This means
that there is $d$ such that 
\begin{numlist}
  \item
  $m\leq e-d$; 
  \item
  $t\leq_{d^2}^n\emptytree$;
  \item
  for every $s\in T$, $s\leq_{d}^n\emptytree$.
\end{numlist} 
If we put $f(i)=\emptytree$ for every $i$, we get 
$t\leq_{d^2}^n\emptytree+\sum_{i=1}^d f(i)$, which
yields $\node{m}{t}{T}\leq_e^{n+1}\emptytreeone$.
In the same way, we can prove that if $\node{m}{t}{T}\leq_e^{n+1}\emptytreeone$,
then $\node{m}{t}{T}\leq_e^{n+1}\emptytree$.\par
We have:
\begin{eqnarray*}
\distpartwo{e}{n+1}{\emptytree}{\emptytree}&=&e+\distpartwo{e^2}{n}{\emptytree}{\emptytree}\\
\distpartwo{e}{n+1}{\emptytree}{\emptytreeone}&=&e+\distpartwo{e^2}{n}{\emptytree}{\emptytree}\\
\distpartwo{e}{n+1}{\emptytreeone}{\emptytree}&=&e+\distpartwo{e^2}{n}{\emptytree}{\emptytree}\\
\distpartwo{e}{n+1}{\emptytree}{\node{m}{t}{T}}&=&m+e+\max_f\{\distpartwo{(m+e)^2}{n}{\emptytree}
  {t+\sum_{i=1}^{m+e} f(i)}\}\\
&=&\distpartwo{e}{n+1}{\emptytreeone}{\node{m}{t}{T}}\\
\distpartwo{e}{n+1}{\node{m}{t}{T}}{\emptytree}&=&e-m+\distpartwo{(e-m)^2}{n}{t}{\emptytree}\\
&=&\distpartwo{e}{n+1}{\node{m}{t}{T}}{\emptytreeone}\\
\end{eqnarray*}
Moreover, observe that 
\begin{eqnarray*}
\emptytree+\emptytree=\emptytree&\cong&\emptytreeone=\emptytreeone+\emptytree\\
\node{m}{t}{T}+\emptytree&=&\node{m}{t}{T}+\emptytreeone
\end{eqnarray*}
This concludes the proof.\qed
\end{proof}
\begin{proposition}[Compatibility]\label{prop:compatibilitylight}
For every $n,e\in\N$, $\emptytree\leq_e^n t$ for every $t$ and, moreover, 
if $t\leq_e^n u$ then $t+v\leq_e^n u+v$ for every $t,u,v$. 
\end{proposition}
\begin{proof}
$\emptytree\leq_e^n t$ is trivial. The second statement
can be proved by induction on $n$. The base case is trivial.
In the inductive case, we can suppose all the involved trees 
to be different from $\emptytree$.
Suppose that $\node{m}{t}{T}\leq^{n+1}_e\node{l}{u}{U}$. 
We should prove $\node{m+k}{t+v}{T\cup V}\leq^{n+1}_e\node{l+k}{u+v}{U\cup V}$.
However,  
\begin{eqnarray*}
m+k&\leq&(l+e)-d+k=(l+k+e)-d\\
t+v&\leq_{d^2}^n&u+\sum_{i=1}^d f(i)+v=u+v+\sum_{i=1}^{d}f(i)\\
\end{eqnarray*}
Moreover, for every $z\in T\cup V$ there certanily 
exists $w\in U\cup V$ such that $z\leq^n_d w$.\qed
\end{proof}
\begin{proposition}[Transitivity]\label{prop:transitivitylight}
If $t \leq_e^n u\leq_d^n v$, then
$t\leq_{d+e}^n v$. 
\end{proposition} 
\begin{proof}
We go by induction on $n$. We can directly go to the
inductive case, since if $n=0$, then the thesis is trivial.
We can assume all the involved trees to be different from $\emptytree$.
Let us suppose $\node{m}{t}{T}\leq_e^{n+1}\node{l}{u}{U}$
and $\node{l}{u}{U}\leq_d^{n+1}\node{k}{v}{V}$
First of all, we have $m\leq l+e-c$ and $l\leq k+d-b$, which
yields $m\leq k+d-b+e-c=k+(d+e)-(b+c)$. Moreover, by hypothesis,
there are functions $f:\{1,\ldots,c\}\rightarrow U$ and
$g:\{1,\ldots,b\}\rightarrow V$ such that
\begin{eqnarray*}
t&\leq_{c^2}^n& u+\sum_{i=1}^c f(i)\\ 
u&\leq_{b^2}^n& v+\sum_{i=1}^b g(i)
\end{eqnarray*}
Therefore, by inductive hypothesis and by proposition~\ref{prop:compatibilitylight}:
\begin{eqnarray*}
t&\leq_{c^2+b^2}^n& v+\sum_{i=1}^c f(i)+\sum_{i=1}^b g(i)\\ 
&\leq_{bc}^n& v+\sum_{i=1}^c h(i)+\sum_{i=1}^b g(i)
\end{eqnarray*}
where $h:\{1,\ldots,c\}\rightarrow V$. We can then
find a function $k:\{1,\ldots,c+b\}\rightarrow V$ such
that
$$
t\leq_{(c+b)^2}^n v+\sum_{i=1}^{c+b}k(i).
$$
Finally, if $z\in T$ then we find $w\in U$ such that $z\leq_c^n w$. We
then find $x\in V$ such that $w\leq_b^n x$ and so $z\leq_{c+b}^n x$.\qed 
\end{proof}
\begin{proposition}\label{prop:compatdifflight}
For every $n,e$ and for every $t,u,v$, 
$\distpartwo{e}{n}{t}{u}\leq\distpartwo{e}{n}{t+v}{u+v}$
\end{proposition}
\begin{proof}
We can proceed by induction on $n$ and, again, the case $n=0$ is trivial.
In the inductive case, as usual, we can suppose all the involved trees to be 
different from $\emptytree$. We have
\begin{eqnarray*}
&&\distpartwo{e}{n+1}{\node{m}{t}{T}}{\node{l}{u}{U}}\\ 
&=& l+e-m+\max_f\{\distpartwo{(l+e-m)^2}{n}{t}{u+\sum_{i=1}^{l+e-m}f(i)}\}\\
&=& l+e-m+\distpartwo{(l+e-m)^2}{n}{t}{u+\sum_{i=1}^{l+e-m}f(i)}\\ 
\end{eqnarray*}
where $f$ and realizes the max. By induction hypothesis,
\begin{eqnarray*}
&&\distpartwo{e}{n+1}{\node{m}{t}{T}}{\node{l}{u}{U}}\\
&\leq& (l+k)+e-(m+k)+\distpartwo{((l+k)+e-(m+k))^2}{n}{t+v}
{u+v+\sum_{i=1}^{(l+k)+e-(m+k)}f(i)}\\
&\leq& \distpartwo{e}{n+1}{\node{m}{t}{T}+\node{k}{v}{V}}{\node{l}{u}{U}+\node{k}{v}{V}}
\end{eqnarray*}
This concludes the proof.\qed
\end{proof}
\begin{proposition}\label{prop:transdifflight}
$\distpartwo{e}{n}{t}{u}+\distpartwo{d}{n}{u}{v}\leq\distpartwo{e+d}{n}{t}{v}$
\end{proposition}
\begin{proof}
We can proceed by induction on $n$ and, again, the case $n=0$ is trivial.
In the inductive case, as usual, we can suppose all the involved trees to be 
different from $\emptytree$. Now
\begin{eqnarray*}
&&\distpartwo{e}{n+1}{\node{m}{t}{T}}{\node{l}{u}{U}}
 +\distpartwo{d}{n+1}{\node{l}{u}{U}}{\node{k}{v}{V}}\\ 
&=&l+e-m+\max_f\{\distpartwo{(l+e-m)^2}{n}{t}{u+\sum_{i=1}^{l+e-m}f(i)}\}\\ 
&&+k+d-l+\max_g\{\distpartwo{(k+d-l)^2}{n}{u}{v+\sum_{i=1}^{k+d-l}g(i)}\}\\
&=&k+(e+d)-m+\distpartwo{(l+e-m)^2}{n}{t}{u+\sum_{i=1}^{l+e-m}f(i)}\\ 
&& +\distpartwo{(k+d-l)^2}{n}{u}{v+\sum_{i=1}^{k+d-l}g(i)}\\
&=&k+(e+d)-m+\distpartwo{(l+e-m)^2}{n}{t}{u+\sum_{i=1}^{l+e-m}f(i)}\\ 
&& +\distpartwo{(k+d-l)^2}{n}{u+\sum_{i=1}^{l+e-m}f(i)}{v+\sum_{i=1}^{k+d-l}g(i)+\sum_{i=1}^{l+e-m}f(i)}\\
&\leq&k+(e+d)-m+\distpartwo{(l+e-m)^2+(k+d-l)^2}{n}{t}{v+\sum_{i=1}^{k+d-l}g(i)+\sum_{i=1}^{l+e-m}f(i)}
\end{eqnarray*}
A function $h:\{1,\ldots,l+e-m\}\rightarrow V$ such that
$\sum_{i=1}^{l+e-m}f(i)\leq^n_{(l+e-m)(k+d-l)}\sum_{i=1}^{l+e-m}h(i)$
can be easily defined, once we remember that 
$\node{l}{u}{U}\leq_d^n\node{k}{v}{V}$. This yields
\begin{eqnarray*}
&&\distpartwo{e}{n+1}{\node{m}{t}{T}}{\node{l}{u}{U}}
 +\distpartwo{d}{n+1}{\node{l}{u}{U}}{\node{k}{v}{V}}\\ 
&\leq&k+(e+d)-m+\distpartwo{(l+e-m)^2+(k+d-l)^2}{n}{t}{v+\sum_{i=1}^{k+d-l}g(i)+\sum_{i=1}^{l+e-m}f(i)}\\
&&+\distpartwo{(l+e-m)(k+d-l)}{n}{v+\sum_{i=1}^{k+d-l}g(i)+\sum_{i=1}^{l+e-m}f(i)}
  {v+\sum_{i=1}^{k+d-l}g(i)+\sum_{i=1}^{l+e-m}h(i)}\\
&\leq&k+(e+d)-m+\distpartwo{(k+(e+d)-m)^2}{n}{t}{v+\sum_{i=1}^{k+d-l}g(i)+\sum_{i=1}^{l+e-m}h(i)}\\
&\leq&k+(e+d)-m+\distpartwo{(k+(e+d)-m)^2}{n}{t}{v+\sum_{i=1}^{l+(d+e)-m}p(i)}
\end{eqnarray*}
where $p:\{1,\ldots,l+(d+e)-m\}\rightarrow V$,
$p(i)=f(i)$ if $i\leq l+e-m$ and $p(i)=g(i-(l+e-m))$
otherwise. But, then
\begin{eqnarray*}
&&\distpartwo{e}{n+1}{\node{m}{t}{T}}{\node{l}{u}{U}}
 +\distpartwo{d}{n+1}{\node{l}{u}{U}}{\node{k}{v}{V}}\\ 
&\leq&\distpartwo{e+d}{n}{\node{m}{t}{T}}{\node{k}{v}{V}}
\end{eqnarray*}
This concludes the proof.\qed
\end{proof}
\begin{lemma}\label{lemma:updepth}
For every $t,u,e$, if $t\leq_e^{\max\{\depth{t},\depth{u}\}}u$,
then for every $n>\max\{\depth{t},\depth{u}\}$, $t\leq_e^n u$
and $\distpartwo{e}{n}{t}{u}=\distpartwo{e}{\max\{\depth{t},\depth{u}\}}{t}{u}$. 
\end{lemma}
\begin{proof}
A straightforward induction on $\max\{\depth{t},\depth{u}\}$.\qed
\end{proof}}{}
The binary relation
$\leq_{\trees}$ on $|\trees|$ is defined by putting $t\leq_{\trees} u$
whenever $\depth{t}\leq\depth{u}$ and $t\leq_0^{\depth{u}} u$.
$\dist_{\trees}$ is defined by 
letting $\distpar{\trees}{t}{u}=\distpartwo{0}{\depth{u}}{t}{u}$.
\begin{lemma}
$\trees=(|\trees|,+,\leq_{\trees},\dist_{\trees})$ is a resource monoid.
\end{lemma} 
\condinc{\begin{proof}
$(|\trees|,+)$ is certainly a commutative monoid. For every $t$,
$t\leq_{\trees}t$, as can be proved by induction on $t$:
$\emptytree\leq_0^0\emptytree$ by definition and, moreover,
$t=\node{m}{u}{U}\leq_0^{\depth{t}} t$ because, by inductive
hypothesis, $u\leq_0^{\depth{u}}u$ which yields, by lemma~\ref{lemma:updepth},
$u\leq_0^{\depth{t}-1}u$. In the same way, we can prove
that, for every $v\in U$, $v\leq_0^{\depth{t}-1}v$. Now, suppose
$t\leq_{\trees}u$ and $u\leq_{\trees}v$. This means that
$t\leq_0^{\depth{u}}u$, $u\leq_0^{\depth{v}}v$,
$\depth{t}\leq\depth{u}$ and $\depth{u}\leq\depth{v}$.
We can then conclude that $\depth{t}\leq\depth{v}$,
that $t\leq_0^{\depth{v}}u$ (by lemma~\ref{lemma:updepth})
and $t\leq_0^{\depth{v}}v$ (by proposition~\ref{prop:transdifflight}).
This in turn yields $t\leq_{\trees}v$. Let us now prove compatibility:
suppose $t\leq_{\trees}u$ and let $v$ be a tree. Then 
$\depth{t}\leq\depth{u}$ and $t\leq_0^{\depth{u}}u$. If
$\depth{v}\leq\depth{u}$, then $\depth{u+v}=\depth{u}$ and
we can proceed by getting $t+v\leq_0^{\depth{u+v}}u+v$
(by proposition~\ref{prop:compatibilitylight}), which means 
$t+v\leq_{\trees}u+v$. If, on the other hand, $\depth{v}>\depth{u}$,
then we can first apply lemma~\ref{lemma:updepth} obtaining
$t\leq_0^{\depth{u+v}}u$ and then $t+v\leq_0^{\depth{u+v}}u+v$
(by proposition~\ref{prop:compatibilitylight}). By way
of lemma~\ref{lemma:updepth} and 
propositions~\ref{prop:transdifflight} and~\ref{prop:compatdifflight}
we get
\begin{eqnarray*}
\distpar{\trees}{t}{u}+\distpar{\trees}{u}{v}&=&
  \distpartwo{0}{\depth{u}}{t}{u}+\distpartwo{0}{\depth{v}}{u}{v}\\
&=&\distpartwo{0}{\depth{v}}{t}{u}+\distpartwo{0}{\depth{v}}{u}{v}\\
&\leq&\distpartwo{0}{\depth{v}}{t}{v}=\distpar{\trees}{t}{v}\\
\distpar{\trees}{t}{u}&=&
  \distpartwo{0}{\depth{u}}{t}{u}\leq\distpartwo{0}{\depth{u+v}}{t}{u}\\
&\leq&\distpartwo{0}{\depth{u+v}}{t+v}{u+v}=\distpar{\trees}{t+v}{u+v}\\
\end{eqnarray*}
This concludes the proof.\qed
\end{proof}}{}
A \emph{light length space} is a length space on the resource monoid
$\trees=(|\trees|,+,\leq_{\trees},\dist_{\trees})$.       
Given a light length space $A=(|A|,\vdashp{A})$, we can define:
\begin{varitemize}
  \item
  The light length space $!A=(|A|,\vdashp{!A})$ where
  $\maj{!A}{t}{e}{a}$
  iff $\maj{A}{u}{e}{a}$ and 
  $t\geq_{\trees}!u$.
  \item
  The light length space $\S A=(|A|,\vdashp{\S A})$ where
  $\maj{\S A}{t}{e}{a}$
  iff $\maj{A}{u}{e}{a}$ and 
  $t\geq_{\trees}\S u$.
\end{varitemize}
The following results states the existence of certain morphisms
and will be useful when interpreting light affine logic.
\begin{lemma}[Basic Maps]
Given light length spaces $A,B$, there are morphisms:
$\mathit{contr}:!A\rightarrow!A\otimes!A$, 
$\mathit{distr}:\S A\otimes\S B\rightarrow \S(A\otimes B)$
and $\mathit{derelict}:!A\rightarrow \S A$ where
$\mathit{contr}(a)=(a,a)$ and $\mathit{distr}(a,b)=(a,b)$
and $\mathit{derelict}(a)=a$.
\end{lemma}
\begin{proof}
We know that $\{e_\mathit{contr}\}(d)$ takes time at most 
$|d|+p$, where $p$ is a constant. Then, let $t,u\in |\trees|$ be
such that $\normpar{\trees}{t}\geq p+|e_\mathit{contr}|$,
$\normpar{\trees}{u}\geq\cp$. Define $t_\mathit{contr}$
to be $t+u+\basictree{2}$. Clearly, $\normpar{\trees}{t_\mathit{contr}}\geq |e_\mathit{contr}|$.
Now, let $\maj{!A}{v}{d}{a}$. This means that
$v\geq_{\trees}!w$ where $\maj{A}{w}{d}{a}$.
Then:
\begin{eqnarray*}
u+!w+!w&\geq_{\trees}& !w+!w\\
\normpar{\trees}{u+!w+!w}&\geq&\normpar{\trees}{u}+\normpar{\trees}{!w}+\normpar{\trees}{!w}\\
&\geq&\cp+\normpar{\trees}{!w}+\normpar{\trees}{!w}\geq |\langle d,d\rangle|
\end{eqnarray*}
This implies $\maj{!A\otimes !A}{u+!w+!w}{|\langle d,d\rangle|}{(a,a)}$.
Moreover, $u+!w+!w=u+!w+\basictree{1}\leq_{\trees}v+t_\mathit{contr}$.
Finally,
\begin{eqnarray*}
\timetm{e_\mathit{contr}}{d}&\leq&|d|+p\leq \normpar{\trees}{w}+\normpar{trees}{t}\\
&\leq& \distpar{\trees}{u+!w+!w}{!w+t_\mathit{contr}}\leq \distpar{\trees}{u+!w+!w}{v+t_\mathit{contr}}
\end{eqnarray*}
This proves $\mathit{contr}$ to be a morphism.\par
Let $e_\mathit{distr}=e_\mathit{id}$. We know that 
$\{e_\mathit{id}\}(d)$ takes constant time, say at
most $p$. Then, let $t,u\in |\trees|$ be
such that $\normpar{\trees}{t}\geq p+|e_\mathit{distr}|$,
$\normpar{\trees}{u}\geq\cp$. $t_\mathit{distr}$ is then
defined as $t+\S u$. 
Now, let $\maj{\S A\otimes \S B}{v}{\langle d,c\rangle}{(a,b)}$. 
This implies that $v\geq \S w+\S x$, where 
$\maj{A}{w}{d}{a}$ and $\maj{B}{x}{c}{b}$. This in turn
means that $\maj{A\otimes B}{w+x+u}{\langle d,c\rangle}{(a,b)}$
and  $\maj{A\otimes B}{\S(w+x+u)}{\langle d,c\rangle}{(a,b)}$.
Moreover
$$
\S(w+x+u)=\S w+\S x+\S u\leq v+t_\mathit{distr}
$$
Finally:
\begin{eqnarray*}
\timetm{e_\mathit{distr}}{\langle d,c\rangle}&\leq&p\leq\normpar{\trees}{t}\\
&\leq& \distpar{\trees}{0}{t}+\distpar{\trees}{\S(w+x+u)}{v+\S u}\leq\distpar{\trees}{\S(w+x+u)}{v+t_\mathit{distr}}
\end{eqnarray*}
This proves $\mathit{distr}$ to be a morphism.\par
Let $e_\mathit{derelict}=e_\mathit{id}$. We know that 
$\{e_\mathit{derelict}\}(d)$ takes constant time,
say at most $p$. Then, let $t_\mathit{distr}\in |\trees|$ be
such that $\normpar{\trees}{t_\mathit{distr}}\geq p+|e_\mathit{derlict}|$. 
Now, let $\maj{!A}{v}{d}{a}$. 
This means that $v\geq !w$, where 
$\maj{A}{w}{d}{a}$. This in turn
means that $\maj{\S A}{\S w}{d}{a}$.
Moreover
$$
\S w\leq !w\leq !w+t_\mathit{derelict}.
$$
Finally:
\begin{eqnarray*}
\timetm{e_\mathit{distr}}{d}&\leq&p\leq\normpar{\trees}{t_\mathit{derelict}}\\
&\leq& \distpar{\trees}{0}{t_\mathit{derelict}}+\distpar{\trees}{\S w}{!w}\\
&\leq& \distpar{\trees}{\S w}{!w+t_\mathit{derelict}}
\end{eqnarray*}
This proves $\mathit{derelict}$ to be a 
morphism.\qed
\end{proof}
\begin{lemma}\label{lemma:lifting}
For every $t\in |\trees|$, there is $u$ such that,
for every $v$, $!(v+t)\leq_{\trees}!v+u$.
\end{lemma}
\condinc{\begin{proof}
First of all we will prove the following statement by induction
on $t$: for every $t$, there is an integer $\overline{t}$ such 
that for every $u$, $u+t\leq_{\overline{t}}^{\max\{\depth{u},\depth{t}\}}u$.
If $t=\emptytree$, we can choose $\overline{t}$ to be just $0$,
since $u\leq_0^n u$ for every $u$. If $t=\node{m}{v}{V}$,
then we put $\overline{t}=m+\overline{v}+\sum_{w\in V}\overline{w}$. 
Let $u$ be an arbitrary tree and let us assume, without losing
generality, that $u=\node{l}{w}{W}$. Let $d=\overline{v}+\sum_{w\in V}\overline{w}$.
We get
\begin{eqnarray*}
l+m&\leq&l+m+(\overline{v}+\sum_{w\in V}\overline{w})
            - (\overline{v}+\sum_{w\in V}\overline{w})\\
&=& l+\overline{t}-d\\
v+w&\leq_{\overline{v}}^{\max\{\depth{v},\depth{w}\}}&w\\
   &\leq_0^{\max\{\depth{v},\depth{w}\}}&w+\sum_{i=1}^d\emptytree\\ 
\forall x\in V. x&\leq_{\overline{x}}^{\depth{x}}&\emptytree\\
\forall x\in W. x&\leq_0^{\depth{x}}&x 
\end{eqnarray*}
Using known results, we can rewrite these inequalities as
follows
\begin{eqnarray*}
l+m&\leq&l+\overline{t}-d\\
v+w&\leq_{d^2}^{\max\{\depth{t},\depth{u}\}-1}&w+\sum_{i=1}^d\emptytree\\ 
\forall x\in V. x&\leq_{d}^{\max\{\depth{t},\depth{u}\}-1}&\emptytree\\
\forall x\in W. x&\leq_{d}^{\max\{\depth{t},\depth{u}\}-1}&x 
\end{eqnarray*}
This yields $u+t\leq_{\overline{t}}^{\max\{\depth{u},\depth{t}\}}t$.\par
Let us now go back to the lemma we are proving. We will now prove that
for every $t$, any term $u=\node{\overline{t}}{w}{U}$ such that 
$\depth{u}=\depth{t}+1$ satisfies the thesis. Indeed, if we
put $d=\overline{t}$ and $n=\depth{v+t}$, we get:
\begin{eqnarray*}
1&\leq&\overline{t}-d+1\\ 
\emptytree&\leq_{d^2}^n& u\\
v+t&\leq_d^n&v
\end{eqnarray*}
This, in turn implies $!(v+t)\leq_0^{n+1}!v+u$, which
yields $!(v+t)\leq_{\trees}!v+u$.\qed 
\end{proof}}{} 
\begin{lemma}[Functoriality]
If $f:\morphism{A}{e}{\varphi}{B}$, then there are $\psi,\theta$ such
that $f:\morphism{!A}{e}{\psi}{!B}$ and $f:\morphism{\S A}{e}{\theta}{\S B}$. 
\end{lemma}
\condinc{\begin{proof}
Let $\xi$ be the tree obtained from $\varphi$ by lemma~\ref{lemma:lifting} and
put $\psi=\xi+\varphi+\basictree{1}$. Suppose that $\maj{!A}{t}{d}{a}$. Then $t\geq !u$,
where $\maj{A}{u}{d}{a}$. Observe that there must be $v,c$ such that
$\maj{B}{v}{c}{f(a)}$, $v\leq_{\trees} u+\varphi$ and
$\timetm{e}{d}\leq\normpar{\trees}{u+\varphi}\distpar{\trees}{v}{u+\varphi}$.
But then $\maj{!B}{!v}{c}{f(a)}$ and moreover
\begin{eqnarray*}
!v&\leq_{\trees}&!(u+\varphi)\leq_{\trees}!u+\xi\leq_{\trees}t+\psi\\
\timetm{e}{d}&\leq&\distpar{\trees}{v}{u+\varphi}\leq\distpar{\trees}{!v}{!(u+\varphi)+\basictree{1}}\\
  &\leq&\distpar{\trees}{!v}{!u+\xi+\basictree{1}}\leq\distpar{\trees}{!v}{t+\psi}
\end{eqnarray*}
This means that $f:\morphism{!A}{e}{\psi}{!B}$. Now, let $\theta$
be $\S\varphi$ and suppose $\maj{\S A}{t}{d}{a}$. Then $t\geq \S u$,
where $\maj{A}{u}{d}{a}$. Observe that there must be $v,c$ such that
$\maj{B}{v}{c}{f(a)}$, $v\leq_{\trees} u+\varphi$ and
$\timetm{e}{d}\leq\normpar{\trees}{u+\varphi}\distpar{\trees}{v}{u+\varphi}$.
But then $\maj{\S B}{\S v}{c}{f(a)}$ and, moreover
\begin{eqnarray*}
\S v&\leq_{\trees}&\S(u+\varphi)=\S u+\S\varphi\leq_{\trees}t+\theta\\
\timetm{e}{d}&\leq&\distpar{\trees}{v}{u+\varphi}\leq\distpar{\trees}{\S v}{\S(u+\varphi)}\\
  &\leq&\distpar{\trees}{\S v}{\S u+\S \varphi)}\leq\distpar{\trees}{\S v}{t+\theta}
\end{eqnarray*}
This means that $f:\morphism{\S A}{e}{\theta}{\S B}$.\qed
\end{proof}}{}
Now, we can prove a polynomial bound on $\normpar{T}{t}$:
\begin{proposition}
For every $n\in\N$ there is a polynomial $p_n:\N\rightarrow\N$ such
that $\normpar{\trees}{t}\leq p_{\depth{t}}(|t|)$.
\end{proposition}
\condinc{\begin{proof}
We prove a stronger statement by induction on $n$: for every $n\in\N$
there is a polynomial $q_n:\N^2\rightarrow\N$ such that for every $t,e$,
$\distpartwo{e}{n}{\emptytree}{t}\leq q_n(|t|,e)$. First of all, we
know that $\distpartwo{e}{0}{\emptytree}{t}=0$, so $q_0$ is just the
function which always returns $0$. $q_{n+1}$ is defined from $q_n$ as
follows: $q_{n+1}(x,y)=x+y+q_n(x(x+y+1),(x+y)^2)$.
Indeed:
\begin{eqnarray*}
\distpartwo{e}{n+1}{\emptytree}{\emptytree}&=&e+\distpartwo{e}{n}{\emptytree}{\emptytree}\\
  &\leq&e+q_n(0,e)\leq e+|\emptytree|\\
  &&+q_n(|\emptytree|(|\emptytree|+e+1),(|\emptytree|+e)^2)\\
  &=&q_{n+1}(|\emptytree|,e)\\
\distpartwo{e}{n+1}{\emptytree}{\node{m}{t}{T}}&=&
     m+e+\max_f\{\distpartwo{(m+e)^2}{n}{\emptytree}{t+\sum_{i=1}^{m+e}f(i)}\}\\
  &\leq&m+e+q_n((m+e+1)(|\node{m}{t}{T}|),(m+e)^2)\\
  &\leq&|\node{m}{t}{T}|+e\\
  &&+q_n((|\node{m}{t}{T}|+e+1)(|\node{m}{t}{T}|),(|\node{m}{t}{T}|+e)^2)\\
  &\leq&q_{n+1}(|\node{m}{t}{T}|,e)
\end{eqnarray*}
At this point, however, it suffices to put $p_n(x)=q_n(x,0)$.\qed
\end{proof}}{}
As for \EAL and \SAL, we cannot claim $(n,m)\mapsto p_n(m)$ to be
a polynomial. However, this is not a problem since we will be
able to majorize binary strings by trees with bounded depth (cf.Remark~\ref{remark:bounddepth}).
\subsection{Interpreting Light Affine Logic}
As for the $!$ modality, $\reasem{\S A}{\eta}=\S\reasem{A}{\eta}$.
\begin{theorem}
Light length spaces form a model of \LAL.
\end{theorem}
Binary lists can be represented in \LAL\ as cut-free proofs
with conclusion
$$
\Listlal\equiv\forall\alpha.!(\alpha\linear\alpha)\linear !(\alpha\linear\alpha)
\linear\S(\alpha\linear\alpha)
$$
\begin{corollary}[Soundness]
Let $\pi$ be an \LAL\ proof with conclusion $\vdash\{!,\S\}^j\Listlal\linear\{!,\S\}^k\Listlal$
and let $f:B\rightarrow B$ be the function induced by $\reasem{\pi}{}$.
Then $f$ is computable in polynomial time.
\end{corollary}}
{
\section{Other Light Logics}\label{sect:oll}
Girard and Lafont have proposed refinements of \EAL, namely Light
Linear Logic (\LLL) and Soft Linear Logic (\SLL), which capture
polynomial time. We have succeeded in defining appropriate reource
monoids for affine variants of these logics, too. In this way 
we can obtain proofs of ``polytime soundness'' by performing the 
same realizability interpretation as was exercised in the previous section. These
instantiations of our framework are considerably more technical and
difficult to find, but share the idea of the \EAL\ interpretation which
is why we have decided not to include them in this Extended
Abstract. The interested reader may consult the full paper (or the
appendix). 

In the following section, we will elaborate in some more detail a
rather different instantiation of our method. 
}
\section{Interpreting \LFPL}\label{sect:lfpl}
In~\cite{hofmann99lics} one of us had introduced another
language, \LFPL, with the property that all definable functions on
natural numbers are polynomial time computable. The key difference
between \LFPL\ and other systems is that a function defined by iteration
or recursion is not marked as such using modalities or similar and can
therefore be used as a step function of subsequent recursive
definitions.

In this section we will describe a resource monoid $\MLFPL$ for \LFPL,
which will provide a proof of polytime soundness for
that system. This is essentially the same as the proof from~\cite{hofmann99lics}, 
but more structured and, hopefully, easier to understand.

The new approach also yields some new results, namely the
justification of second-order quantification, a !-modality, and a new
type of binary trees based on cartesian product which allows
alternative but not simultaneous access to subtrees.

\subsection{Overview of \LFPL}
\LFPL\ is intuitionistic, affine linear logic, i.e., a linear functional
language with $\otimes, \linear, +, \times$. Unlike in the original
presentation we also add polymorphic quantification here. In addition,
\LFPL\ has basic types for inductive datatypes, for example unary and
binary natural numbers, lists, and trees. There is one more basic
type, namely $\Diamond$, the resource type. 

The recursive constructors for the inductive datatypes each take an additional
argument of type $\Diamond$ which prevents one to invoke more
constructor functions than one.
Dually to the constructors one has iteration principles
which make the $\Diamond$-resource available in the branches of a
recursive definition. For example, the type $T(X)$ of $X$-labelled
binary trees has constructors $\mathbf{leaf}:T(X)$ and
$\mathbf{node}:\Diamond\linear X\linear T(X)\linear T(X)\linear
T(X)$. The iteration principle allows one to define a function
$T(X)\linear A$ from closed terms  $A$ and
$\Diamond\linear X\linear A\linear A\linear A$. 

In this paper we ``internalise''  the assumption of closedness using a
$!$-modality. 

Using this iteration principle one can encode recursive definitions by
ML-style pattern matching provided recursive calls are made on
structurally smaller arguments only. 

Here is a fragment of an \LFPL\ program for ``treesort'' written in
functional notation: the additional arguments of type $\Diamond$ are
supplied using @. Note that the insert function takes an extra
argument of type $\Diamond$. 
{\small
\begin{verbatim}
let insert x t d = match t with 
   Leaf -> Node(x,Leaf,Leaf)@d
 | Node(y,l,r)@d' -> 
  if x<=y then Node(y,insert x l d,r)@d' 
          else Node(y,l,insert x r d)@d'

let extract t = match t with 
   Leaf -> nil
 | Node(x,l,r)@d -> 
  append (extract l) (cons(x,extract r)@d) 
\end{verbatim}}

\subsection{A Resource Monoid for \LFPL}
The underlying set of $\MLFPL$ is the set of pairs $(l,p)$ where
$l\in\mathbb{N}$ is a natural number and $p$ is a monotone polynomial
in a single variable $x$. The addition is defined by
$(l_1,p_1)+(l_2,p_2)=(l_1+l_2,p_1+p_2)$, accordingly, the neutral
element is $0=(0,0)$. We have a submonoid $\MLFPLZ=\{(l,p)\in
\MLFPL\mid l=0\}$.

To define the ordering we set $(l_1,p_1)\leq(l_2,p_2)$ iff $l_1\leq
l_2$ and $(p_2-p_1)(x)$ is monotone and nonnegative for all $x\geq
l_2$. For example, we have $(1,42x)\leq (42,x^2)$, but
$(1,42x)\not\leq (41,x^2)$. The distance function is defined by
\[
\distpar{\MLFPL}{(l_1,p_1)}{(l_2,p_2)}=(p_2-p_1)(l_2)
\]
We can pad elements of $\MLFPL$ by adding a constant to the
polynomial. The following is now obvious.
\begin{lemma}
Both $\MLFPL$ and $\MLFPLZ$ are resource monoids.
\end{lemma}
A simple inspection of the proofs in Section~\ref{bloed} shows that
the realisers for all maps can be chosen from $\MLFPLZ$. This is
actually the case for an arbitrary submonoid of a resource monoid.  We
note that realisers of elements may nevertheless be drawn from all of
$\MLFPL$. We are thus led to the following definition.
\begin{definition}
  An \LFPL-space is a length space over the resource monoid $\MLFPL$. A
  morphism from \LFPL\ length space $A$ to $B$ is a morphism between
  length spaces which admits a majorizer from $\MLFPLZ$.
\end{definition}
\begin{proposition}
\LFPL\ length spaces with their maps form a symmetric monoidal 
closed category.
\end{proposition}
\begin{definition}
Let $A$ be an \LFPL\ space and $n\in\mathbb{N}$. The \LFPL\ space $A^n$ is
defined by $|A^n|=|A|$ and $\maj{A^n}{\alpha}{e}{a}$ iff $\alpha\geq
(2n-1).\beta$ for some $\beta$ such that $\maj{A}{\beta}{e}{a}$. 
\end{definition}
So, $A^n$ corresponds to the subset of $A\otimes\dots\otimes A$
consisting of those tuples with all $n$ components equal to each
other. The factor $2n-1$ (``modified difference'') instead of just $n$ is needed in order to justify the linear time needed to compute the copying involved in the obvious morphism from $A^{m+n}$ to $A^m\otimes A^n$. 

Let $I$ be an index set and $A_i, B_i$ be $I$-indexed families of \LFPL\
spaces. A uniform map from $(A_i)_i$ to $(B_i)_i$ consists of a family
of maps $f_i :A_i\rightarrow B_i$ such that there exist $e,\alpha$
with the property that $\maj{}{\alpha}{e}{f_i}$ for all $i$. Recall
that, in particular, the denotations of proofs with free type
variables are uniform maps. 

\begin{proposition} For each $A$ there is a  uniform (in $m,n$) map $A^{m+n}\rightarrow
A^m\otimes A^n$. Moreover, $A^1$ is isomorphic to $A$.
\end{proposition}
The \LFPL-space $\Diamond$ is defined by $|\Diamond| = \{\Diamond\}$ and
put $\maj{\Diamond}{\alpha}{d}{\Diamond}$ if $\alpha\geq (1,0)$. 

For each \LFPL-space $A$ we define \LFPL-space $!A$ by  $|!A|=|A|$ and $\maj{!A}{\alpha}{t}{a}$ if there exists
$\alpha'=(0,p)\in\MLFPLZ$ with $\maj{A}{\alpha'}{t}{a}$ and
$\alpha\geq (0,(x+1)p)$.

\begin{proposition}\label{bangprop}
  There is an \LFPL\ space $\Diamond$ and for each \LFPL\ space $A$ there
  is an \LFPL\ space $!A$ with the following properties:
\begin{varitemize}
\item $|!A| = |A|$.
\item If $f:A\rightarrow B$ then $f:!A\rightarrow!B$. 
\item $!(A\otimes B) \simeq !A\otimes !B$
\item The obvious functions
$!A\otimes \Diamond^{n}\rightarrow A^{n}\otimes
\Diamond^{n}$ are a uniform map. 
\end{varitemize}
The last property means intuitively that with $n$ ``diamonds'' we can
extract $n$ copies from an element of type $!A$ and get the $n$
``diamonds'' back for later use. 
\end{proposition}
\condinc{\begin{proof}
We have $(0+1)p(0)=p(0)\geq |t|$. Compatibility with $\otimes$ is obvious. 

For functoriality assume that $\maj{}{\phi}{e}{f}$ where
$\phi=(0,q)\in\MLFPLZ$. We claim that $\maj{}{(0,(x+1)q)}{e}{f}$
\emph{qua} morphism from $!A$ to $!B$. Suppose that
$\maj{!A}{\alpha}{t}{a}$ where $\alpha\geq (0,(x+1)p)$ and
$\maj{A}{(0,p)}{t}{a}$. Since $f$ is a morphism, we obtain $v,\beta$ such
that $\maj{B}{\beta}{v}{f(a)}$ and $\beta\leq \phi+(0,p)$. This
implies that $\beta\in\MLFPLZ$ as well, say, $\beta=(0,r)$ where
$r\leq p+q$. We also know that $r(0)\geq |v|$ by the definition of
length spaces. Now $\maj{!B}{(0,(x+1)r)}{v}{f(b)}$. On the other hand
$(x+1)r\leq (x+1)(p+q)$. The resource bounds are obvious.

Finally, consider the required morphism $!A\otimes
\Diamond^{n}\rightarrow A^{n}\otimes \Diamond^{n}$. 
Clearly, it may be realised by the identity; we claim that $0$
can serve as a majoriser.  Indeed, a
majoriser of $(a,d)\in |!A \otimes \Diamond^{n}|$ is of
the form $(2n-1,(x+1)p)$ where $(0,p)$ majorises $a$ in $A$. Now,
$(2n-1,(2n-1)p)$ is a majoriser of $(a,d)$ in $A^n\otimes \Diamond^n$. But
$((x+1)-(2n-1)p$ is monotone and nonnegative above $2n-1$. \qed
\end{proof}}{The proof of the last assertion relies on the fact that $(2n-1,(2n-1)p)\leq (2n-1,(x+1)p)$ for arbitrary $n$.}
\paragraph{Remark}
  We remark at this point that we obtain an alternative resource
  monoid $\mathcal{M}_S$ for \SAL\ whose underlying set and ordering are as in
  $\MLFPL$, but whose addition is given by addition as
  $(l_1,p_1)+(l_2,p_2)=(\max(l_1,l_2),p_1+p_2)$. Length spaces over
  $\mathcal{M}_S$ with maps majorised by $\mathcal{M}_S$ 
  (not $\MLFPLZ$) then also form a
  sound model of \SAL. This points to a close relationship between
  \LFPL\ and \SAL\ and also shows a certain tradeoff between the two
  systems. The slightly more complex model is needed for  \LFPL\
  since in \LFPL\ the C-rule  of \SAL\ is so to say internalised in the form
  of the uniform map $!A\otimes \Diamond^n\rightarrow
  A^n\otimes\Diamond^n$. Notice that \SAL's map $!A\rightarrow A^n$
  cannot be uniform. This uniformity of \LFPL\ allows for an internal
  implementation of datatypes and recursion as we now show. 
\begin{definition}
Let $T_i$ be a family of \LFPL\ spaces such that $|T_i| = T$ independent
of $i$. The \LFPL\ space $\exists i.T_i$ is defined by $|\exists
i.T_i|=|T|$ and $\maj{\exists i.T_i}{\alpha}{e}{t}$ if
$\maj{T_i}{\alpha}{e}{t}$ for some $i$. 
\end{definition}
Note that if we have a uniform family of maps $T_i\rightarrow U$ where
$U$ does not depend on $i$ then we obtain a map $\exists i.T_i
\rightarrow U$ (existential elimination). 

Conversely, if we have a uniform family of maps $U_i\rightarrow
V_{f(i)}$ then we get a uniform family of maps $U_i\rightarrow \exists
j.V_j$ (existential introduction).  We will use an informal ``internal
language'' to denote uniform maps which when formalised would amount
to an extension of \LFPL\ with indexed type dependency in the style of
Dependent ML \cite{xi99popl}.
\subsection{Inductive Datatypes}
In order to interpret unary natural numbers, we define $N = \exists
n.N_n$ where 
\[
N_n = \Diamond^{n}\otimes \forall A.(A\linear A)^{n}\linear A\linear A
\]
We can internally define a successor map $\Diamond\otimes N_n\rightarrow
N_{n+1}$ as follows: starting from $d:\Diamond, \vec
d:\Diamond^{n}$ and $f:\forall (A\linear A)^{n}\linear A\linear A$ 
we obtain a member of $\Diamond^{n+1}$ (from $d$ and $\vec d$) and we
define $f':\forall (A\linear A)^{n+1}\linear A\linear A$ as $\lambda
(u^{A{\linear}A},\vec u^{(A{\linear}A)^{n}}).\lambda z^A.u(f\ \vec u\
z)$. From this, we obtain a map $\Diamond\otimes N\rightarrow N$ by
existential introduction and elimination. 

Of course, we also have a constant zero $I\rightarrow N_0$ yielding a
map $I\rightarrow N$ by existential introduction. 

Finally, we can define an iteration map
\[
!(\Diamond\otimes A \linear A) \linear N_n\linear A\linear A
\]
as follows: 
Given $t: !(\Diamond\otimes A\linear A)$ and $(\vec
d,f)\in N_n$ we unpack $t$ using Proposition~\ref{bangprop} to
yield $t'\in ((\Diamond\otimes A)\linear A)^n$ as well as
$\vec d\in\Diamond^{n}$. Feeding these ``diamonds'' one by one
to the components of $t'$ we obtain $t''\in (A\linear A)^{\otimes n}$.
But then $f\ t''$ yields the required element of $A\linear A$. 

Existential elimination now yields a  single map 
\[
!(\Diamond\otimes A \linear A) \linear N\linear A\linear A
\]
\condinc{Similarly, we can interpret binary $X$-labelled trees using a type
family 
\[
T_n = \Diamond^{n}\otimes \forall (X\linear A \linear 
A\linear A)^{n} \linear A^{n+1}\linear A
\]
and defining trees proper as $\exists n.T_{n}$.  We get maps
$\mathbf{leaf}:T_{0}$ and $\mathbf{node}:\Diamond\otimes X\otimes
T_{n_1}\otimes T_{n_2}\rightarrow T_{n_1+n_2+1}$ and an analogous
iteration construct.

Finally, and this goes beyond what was already known, we can define
``lazy trees'' using cartesian product (also known as additive
conjunction).

First, we recall from ordinary affine linear logic that an additive
conjunction can be defined as
\[
A \times B = \forall C.(C\linear A)\otimes (C\linear B) \otimes C
\]
The first projection map $A\times B\rightarrow A$ is given internally
by $\lambda (f^{C{\linear}A},g^{C{\linear}B},c^C).f\ c$. Analogously,
we have a second projection. Given maps $f:C\rightarrow A$ and $g:
C\rightarrow B$ we obtain a map $\langle f,g\rangle : C\rightarrow
A\times B$ internally as $\lambda c^C.(f,g,c)$.

Now, following the pattern of the binary trees $T_{m,n}$ above, we
define another family
\[
T_{d}^{\times} = \Diamond^{d}\otimes \forall A.(X\linear
(A\times A)\linear A)^{d} \linear A\linear A
\]
and $T^{\times}=\exists d.T^{\times}_{d}$.  We get
maps $\mathbf{leaf}:\Diamond\rightarrow T^{\times}_{0}$ 
 and $\mathbf{node}:\Diamond\otimes X\otimes (T_{d_1}\times
T_{d_2})\rightarrow T_{1+\max(d_1,d_2)}$ as well as an analogous
iteration construct.

We describe in detail the construction of the ``node'' map which is
not entirely straightforward. First, we note that for any length
spaces $A, B$ and $m,n$ the obvious map $(\Diamond^m\otimes A)\times
(\Diamond^n\otimes B)\rightarrow \Diamond^{\max(m,n)}\otimes (A\times
B)$ is a morphism. This is because a majoriser of an element of
$(\Diamond^m\otimes A)\times (\Diamond^n\otimes B)$ must be of the
form $(k,p)$ where $k\geq\max(m,n)$ in view of the existence of the
projection maps.

Now suppose we are given (internally) $d:\Diamond, x:X,
\mathit{lr}:T^\times _{d_1}\times T^\times_{d_2}$. Using the just
described morphism we decompose $\mathit{lr}$ into $\vec
d:\Diamond^{\max(d_1,d_2)}$ and $\mathit{lr}': W_{d_1}\times W_{d_2}$
where $W_{i}=(X\linear (A\times A)\linear A)^i \linear A\linear A$. We
have stripped off the universal quantifier.

Now $d$ and $\vec d$ together yield an element of
$\Diamond^{1+\max(d_1,d_2)}$. It remains to construct a member of
$W_{1+\max(d_1,d_2)}$. To this end, we assume $u:X\linear (A\times
A)\linear A$ and $f:(X\linear (A\times A)\linear A)^{\max(d_1,d_2)}$
and define the required element of $A$ as $u\ x\ \langle
\mathit{lr}'.1\ f\ a, \mathit{lr}'.2\ f\ a\rangle$. Here $.1$ and $.2$
denote the projections from the cartesian product. The sharing of the
variables $f$, $a$, $\mathit{lr}'$ is legal in the two components of a
cartesian pairing, but would of course not be acceptable in a
$\tensor$ pairing. We have elided the obvious coercions from
$(\_)^{\max(d_1,d_2)}$ to $(\_)^{d_i}$.

We remark that these cartesian trees are governed by their depth
rather than their number of nodes. We also note that if $X=I$ we can
form the function $\lambda d^{\Diamond}.\lambda
t^{T^\times}.\mathbf{node}\ d\ ()\ \langle t,r\rangle :
\Diamond\linear T^\times\linear T^\times$. Iterating this map yields a
function $N\linear T^\times$ computing full binary trees of a given
depth. Of course, on the level of the realisers, such a tree is not
laid out in full as this would require exponential space, but computed
lazily as subtrees are being accessed.  Exploring the implications of
this for programming is left to future work.}
{In the appendix we also show how to interpret two different kinds of
binary trees.}
\section{Conclusion}
We have given a unified semantic framework with which to establish
soundness of various systems for capturing complexity classes by logic
and programming. Most notably, our framework has all of second-order
multiplicative linear logic built in, so that only the connectives and
modalities going beyond this need to be verified explicitly. 

While resulting in a considerable simplification of previous soundness
proofs, in particular for \LFPL\ and \LAL, our method has also lead to
new results, in particular polymorphism and a modality for \LFPL. 

The method proceeds by assiging both abstract resource bounds in the
form of elements from a resource monoid and resource-bounded
computations to proofs (respectively, programs). In this way, our method can
be seen as a combination of traditional Kleene-style realisability
(which only assigns computations) and polynomial and quasi
interpretation known from term rewriting (which only assigns resource
bounds). An altogether new aspect is the introduction of more general
notions of resource bounds than just numbers or polynomials as
formalised in the concept of resource monoid.  We thus believe that
our methods can also be used to generalise polynomial interpretations
to (linear) higher-order.  

\bibliographystyle{plain}

\end{document}